\documentclass[preprint, aps,prl,onecolumn,showpacs,amsmath,amssymb,groupedaddress,superscriptaddress,12pt]{revtex4-1} 

\usepackage{graphicx} 
\usepackage{dcolumn} 
\usepackage{bm} 
\usepackage{amssymb} 

\usepackage[T1]{fontenc}
\usepackage{amsmath}
\usepackage{float}
\usepackage{color, soul}

\usepackage{setspace}

\makeatletter
\newcommand*{\rom}[1]{\expandafter\@slowromancap\romannumeral #1@}
\makeatother

\makeatletter
\@ifundefined{textcolor}{}
{%
 \definecolor{BLACK}{gray}{0}
 \definecolor{WHITE}{gray}{1}
 \definecolor{RED}{rgb}{1,0,0}
 \definecolor{GREEN}{rgb}{0,1,0}
 \definecolor{BLUE}{rgb}{0,0,1}
 \definecolor{CYAN}{cmyk}{1,0,0,0}
 \definecolor{MAGENTA}{cmyk}{0,1,0,0}
 \definecolor{YELLOW}{cmyk}{0,0,1,0}
}
\makeatother


\begin{document}
\title{Structural-elastic determination of the mechanical lifetime of biomolecules}

\author{Shiwen Guo}
\thanks{The first two authors contributed equally to this work}
\affiliation{Mechanobiology Institute, National University of Singapore, Singapore 117411}

\author{Qingnan Tang}
\thanks{The first two authors contributed equally to this work}
\affiliation{Department of Physics, National University of Singapore, Singapore 117542}

\author{Mingxi Yao}
\affiliation{Mechanobiology Institute, National University of Singapore, Singapore 117411}
\author{Shimin Le}
\affiliation{Department of Physics, National University of Singapore, Singapore 117542}
\author{Hu Chen}
\affiliation{Department of Physics, Xiamen University, Xiamen, China 361005}

\author{Jie Yan}
\email{phyyj@nus.edu.sg}
\affiliation{Mechanobiology Institute, National University of Singapore, Singapore 117411}
\affiliation{Department of Physics, National University of Singapore, Singapore 117542}
\affiliation{Centre for Bioimaging Sciences, National University of Singapore, Singapore 117546}

\date{\today}

\begin{abstract}
The lifetime of protein domains and ligand-receptor complexes under force is crucial for mechanosensitive functions, while many aspects of how force affects the lifetime still remain poorly understood. Here, we report a new analytical expression of the force-dependent molecular lifetime to understand transitions overcoming a single barrier. Unlike previous models derived in the framework of Kramers theory that requires a presumed one-dimensional free energy landscape, our model is derived based on the structural-elastic properties of molecules which is not restricted by the shape and dimensionality of the underlying free energy landscape. Importantly, the parameters of this model provide direct information of the structural-elastic features of the molecules between the transition and the native states. We demonstrate the applications of this model by applying it to explain complex force-dependent lifetime data for several molecules reported in recent experiments, and predict the structural-elastic properties of the transition states of these molecules.
\end{abstract}

\maketitle


\section{Introduction}
It has been known that single cells can sense mechanical properties of their micro-environment, and transduce the mechanical cues into biochemical reactions that eventually affect cell shape, migration, survival, and differentiation \cite{iskratsch2014}. This mechanotransduction requires transmission of force through a number of mechanical linkages, each of which is often composed of multiple linearly arranged force-bearing proteins that are non-covalently linked to one another.  Under force, the domains in each protein in the linkage may undergo mechanical unfolding. In addition, two neighbouring proteins in the linkage can dissociate. Therefore, the force-dependent lifetime of the protein domains and protein-protein complexes is a key factor that affects the mechanotransduction on a particular mechanical linkage. Determining the force-dependent rate (the reciprocal of lifetime) of rupturing a ligand-receptor complex or unfolding a protein domain has been a focus of experimental measurements \cite{Yuan-Chen2017, Marshall-Zhu2003, Jagannathan2012, rakshit2012ideal, Chen2015, edwards2015optimizing}  and theoretical modelling \cite{Bell1978, hummer2003kinetics, Evans1997, dudko2006, cossio2016kinetic, pereverzev2005twopath, barsegov-Thirumalai2005dynamics, Evans-Zhu2004, pierse2017distinguishing, bartolo2002, rebane2016structure}. Previous single-molecule force spectroscopy measurements have revealed complex kinetics for a variety of molecules \cite{Yuan-Chen2017, Marshall-Zhu2003, Jagannathan2012, rakshit2012ideal, Chen2015},  yet the mechanisms still remain elusive. 

An extensively applied phenomenological expression of $k(F)$ was proposed by Bell et al.\cite{Bell1978}: $k(F) = k_\text{0} e^{\beta F \delta^\text{*}}$ , where $\beta=(k_BT)^\text{-1}$, $k_\text{0}$ is the rate in the absence of force and $\delta^*$ is the constant transition distance. This model assumes that the force applied to the molecule results in change of the energy barrier by the amount of $-F\delta^\text{*}$, while the physical basis of this assumption is weak. The limitation of Bell's model has been revealed in many recent experiments that reported complex deviations from its predictions \cite{Yuan-Chen2017, Marshall-Zhu2003, Jagannathan2012, rakshit2012ideal, Chen2015}. 

In order to explain such deviations, several analytical expressions of $k(F)$ were derived based on extending the Brownian dynamics theory from Kramers \cite{kramers1940brownian} for force-dependent dissociation of bonds \cite{hummer2003kinetics, Evans1997, dudko2006, cossio2016kinetic}. The Kramers theory was originally proposed to study kinetics of particle escaping from an energy well through diffusion on a presumed one-dimensional free energy landscape. The theory showed that for sufficiently high barrier, the escaping rate exponentially decreases with the height of the barrier, which proves Arrhenius law for the one dimensional case. In order to derive $k(F)$ in the framework of Kramers theory, the force-dependent unfolding/dissociation kinetics was paralleled to the kinetics of particle escaping from the energy well when the particle is subject to force. Specifically, a fixed zero force free energy landscape $U_\text{0}(x)$ is assumed, which can be described as a function of the molecular extension $x$ along the pulling direction. Under force, the free energy landscape becomes $U(x)=U_\text{0}(x)-Fx$. Assuming a sufficiently high energy barrier such that the energy well and the barrier are well separated and for the cases where $U_\text{0}(x)$ can be approximated by a cusp or a linear-cubic function, an analytical expression of the force-dependent transition rate $k(F)$ was derived by Dudko et al. \cite{dudko2006}, which has been extensively applied to explain experimental data. 

In general, the applications of the expression of $k(F)$ derived based on the framework of Kramers theory are limited by two factors, namely 1) the transition pathway that is one dimensional with the molecular extension as good reaction coordinate, and 2) the shape of the presumed free energy surface $U_\text{0}(x)$. A more recent publication \cite{dudko2008theory} shows that $k(F)$ can be re-expressed as $k(F)=k_\text{0}e^{\beta \int\limits_0^F \delta^*(F')dF'}$ in the framework of Kramers theory, where $\delta^*(F)$ is the average extension difference of the molecule between the transition state and the native state. This expression does not have an explicit dependence on a presumed free-energy landscape $U_\text{0}(x)$. However, in order to actually apply this formula, a presumed one-dimensional free energy landscape is still needed to calculate $\delta^*(F)$. As a result, its application is still limited by these two factors mentioned above. Due to these limitations, although $k(F)$ derived in the framework of Kramers theory can explain mild deviations from Bell's model \cite{Evans1997, dudko2006, cossio2016kinetic}, they typically predict monotonic $k(F)$ and fail to explain more complex experimentally observed kinetics, such as the non-monotonic $k(F)$ reported in several recent experiments \cite{Yuan-Chen2017, Marshall-Zhu2003, rakshit2012ideal}.

Previously, non-monotonic $k(F)$ were typically explained by high-dimension phenomenological models involving multiple competitive pathways or force-dependent selection of multiple native conformations that have access to different pathways \cite{pereverzev2005twopath, barsegov-Thirumalai2005dynamics, Evans-Zhu2004, pierse2017distinguishing, bartolo2002}. For example, the transition rate described by two competitive transition pathways, $k(F)=k_1(F)+k_2(F)$ , each following Bell's model, can explain non-monotonic $k(F)$ with one of the transition distances being negative \cite{pereverzev2005twopath}. On the other hand, models based on force-dependent selection of multiple native conformations that have access to different pathways are much more complex and lack of analytical simplicity for general cases \cite{barsegov-Thirumalai2005dynamics}. Simplification of such models must require additional assumptions on the force-dependence of the selection of native conformations \cite{barsegov-Thirumalai2005dynamics, Evans-Zhu2004, pierse2017distinguishing}. A limitation of all these models is that the model parameters do not provide insights into the structural and physical properties of the molecules in the native and transition states.

We recently reported that $k(F)$ of mechanical unfolding of titin I27 immunoglobulin (Ig) domain exhibits an unexpected ``catch-to-slip'' behaviour at  low force range \cite{Yuan-Chen2017}. It switches from a decreasing function (i.e., ``catch-bond'' behaviour) at forces below 22 pN to an increasing function (i.e., ``slip-bond'' behaviour) at forces greater than 22 pN. The transition state of the titin I27 domain is known to involve a peeled A-A$^\prime$ peptide containing 13 residues \cite{Lu1998, Lu1999, Best2003, Williams2003}. Taking the advantage of the known structures of I27 in its native and the transition states, we analysed the effects of the structural-elastic properties of I27 on its force-dependent unfolding kinetics by applying Arrhenius law. We demonstrated that the entropic elasticity of titin I27 in the two states is responsible for the observed ``catch-to-slip'' behaviour of $k(F)$. Besides suggesting the structural-elastic property of a molecule as a critical factor affecting the force-dependent transition rate, the result also points to a possibility of deriving $k(F)$ based on the structural-elastic properties of molecules in the framework of Arrhenius law. As the derivation of $k(F)$ based on Arrhenius law does not depend on any presumed free energy scape, it is not limited by the dimensionality of the system and the choice of the transition coordinate. Therefore, it is promising to be applied to a broader scope of experimental cases.  

The result described in our previous work \cite{Yuan-Chen2017} is obtained based on the prior knowledge of the structural-elastic properties of I27 in the native and transition states. Unfortunately, such prior knowledge is unavailable for most of other molecules. In order to interpret the force-dependent transition rate based on the structural-elastic properties of molecules for generic cases, it is necessary to derive an expression of $k(F)$ that contains parameters related to the potential structural-elastic properties of the molecule based on Arrhenius law. If this can be achieved, fitting the experimental data using the derived $k(F)$ not only can be applied to explain experimental data but also can provide important insights into the structural-elastic properties of the molecule based on the best-fitting values of the model parameters. To our knowledge, $k(F)$ with such capability has not been derived before.

\section{Results}
\subsection{Deriving $k(F)$ based on the structural-elastic properties of molecules}
In this work, we derived an analytical expression of $k(F)$ on the basis of the structural-elastic features of molecules by applying Arrhenius law. In our derivation, the native state is modelled as a deformable folded structure with a relaxed length $b^\text{0}$, which is the linear distance between the two force-attaching points on the relaxed folded structure, as depicted in Fig. 1A. Here a deformable folded structure refers to that the structure has a certain Yong's modulus, and it can be slightly deformed along the force direction without causing local structural changes. One example is the B-form DNA, which can be extended beyond its relaxed contour length without breaking any Watson-Crick basepairs at forces in 20-40 pN \cite{bouchiat1999estimating, cong2015revisiting}.

\begin{figure}[htbp]
\centering
\includegraphics[width=0.6\textwidth]{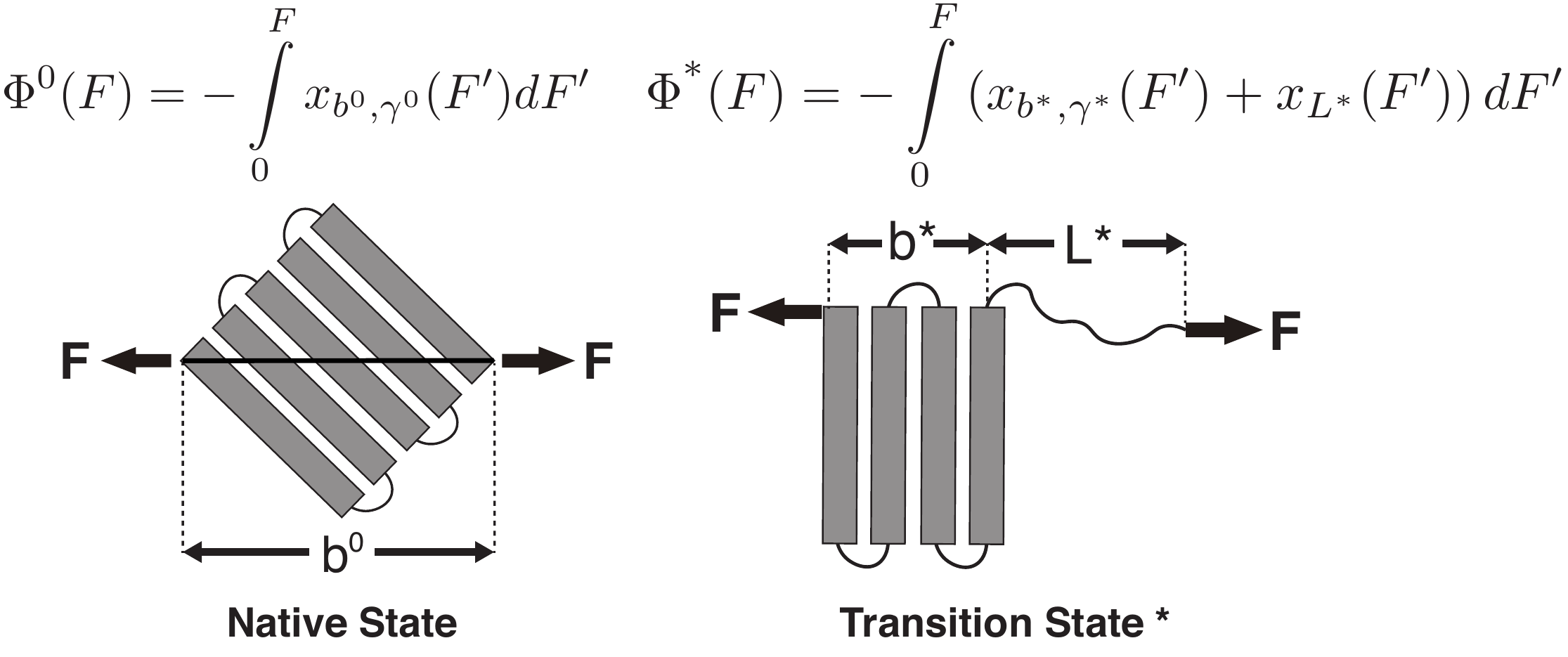}
\caption{{\bf{The entropic elastic free energies of the native and the transition states. }} The native state is sketched as a folded structure, with a length of $b^\text{0}$ and stiffness of $\gamma^\text{0}$. The transition state is modelled as a structure consisting of a folded core with a length of $b^\text{*}$ and stiffness of $\gamma^\text{*}$ as well as a flexible polymer with a contour length of $L^\text{*}$. The force-dependent entropic free energies of the states are indicated.}
\label{Fig.1}
\end{figure}

A deformable folded structure has a very simple analytical force-extension curve \cite{Smith1996}, $x_{b^\text{0},\gamma^\text{0}}(F)=b^\text{0} (\coth({Fb^\text{0} \over k_BT})-{k_BT\over Fb^\text{0}})(1+{F\over \gamma^\text{0}})$, where $b^\text{0}(\coth({Fb^\text{0} \over k_BT})-{k_BT\over Fb^\text{0}})$ is the solution of the force-extension curve of an inextensible rod with a length $b^\text{0}$, and the factor $(1+{F\over \gamma^\text{0}})$ takes into account the force-dependent elongation of the rod. Here $\gamma^\text{0}$ describes the stiffness of the folded structure along the force direction. The ratio $b^\text{0}/\gamma^\text{0}$ describes the deformability of the folded structure. $\gamma$ is in the order of $10^2-10^3$ pN for typical protein domains and nucleic acids structures (SI: S\rom{1}-\rom{2}, Tab. S1).

The transition state is assumed to consist of a deformable folded structure and a semi-flexible polymer that is a peptide for protein or a single-stranded DNA (ssDNA) for DNA structure or a single-stranded RNA (ssRNA) for RNA structure. Its force-extension curve is a sum of the contributions from the folded core and the flexible polymer: $x^\text{*}(F)=x_{b^\text{*}, \gamma^\text{*}}(F)+ x_{L^*}(F)$. Here $b^\text{*}$ and $\gamma^\text{*}$ are the length and stiffness of the folded core, respectively. $L^*=n^* l_\text{r}$ is the contour length of the flexible polymer that contains $n^*$ residues, where $l_\text{r}$ is the contour length per residue.  $x_{L^*}(F)$ can be described by the worm-like chain (WLC) polymer model that contains two parameters, the bending persistence length $A$ and the contour length $L^*=n^* l_\text{r}$. In previous single-molecule manipulation experiments, values of $A\sim 0.8$ nm and $l_\text{r}\sim 0.38$ nm for peptide chain have been experimentally determined \cite{Winardhi2016}. Based on the WLC model, $x_{L^*}(F)$ can be analytically approximated by the Marko-Siggia formula \cite{Marko1995}: ${FA\over k_BT}={x\over L^*}+{1\over 4(1-x/L^*)^2}-{1\over 4}$. 

Force $F$ introduces an entropic  conformational free energy $\Phi(F)$ to a molecule in a particular structural state, in addition to other chemical interactions that maintain the molecule in the structural state. $\Phi(F)$ can be calculated based on the force-extension curve of the molecule as: $\Phi(F)=\int\limits_{0}^{x(F)} f(x') dx'-Fx(F)$, where $f(x)$ is the equilibrium force-extension curve of the molecule, and $x(F)$ is the equilibrium extension of the molecule at the applied force. It is straightforward to show that this energy can be rewritten into a simpler form: $\Phi(F)=-\int\limits_{0}^F x(f') df'$ \cite{cocco2004overstretching, rouzina2001force} (SI: S\rom{3}), where $x(f)$  is the inverse function of $f(x)$. If the molecule has different force-extension curves between the native and the transition states, force applied to the molecule will result in a force-dependent transition distance:
\begin{equation}\label{Eq_td}
\delta^*(F)=x_{b^\text{*}, \gamma^*}(F)-x_{b^\text{0}, \gamma^\text{0}}(F)+ x_{{L^*}}(F),
\end{equation}
and cause a change in the transition free energy barrier that can be computed as: $\Delta \Phi^*(F)=-\int\limits_{0}^F \delta^\text{*}(F') dF'$.

$\Delta \Phi^*(F)$ can be rewritten as a linear combination of three terms: 
\begin{equation}\label{Eq_DPhi}
\Delta \Phi^*(F)=\Phi_{b^*, \gamma^*}(F)-\Phi_{b^\text{0},\gamma^\text{0}}(F)+ \Phi_{{L^*}}(F),
\end{equation} 
where $\Phi_{i}(F)=-\int\limits_{0}^F x_{i}(F') dF'$ denotes the contributions from the folded native state ($_{b^\text{0},\gamma^{\text{0}}}$), the folded core of the transition state ($_{b^*,\gamma^*}$), and the flexible polymer in the transition state ($_{{L^*}}$), respectively. The three force-dependent entropic conformational free energy terms scaled by $\beta^{-1}=k_\text{B}T$ have the following analytical solutions:

\begin{equation}
\label{Eq_Phi}
\left\{
\begin{array}{l}

 \begin{array}{ll}
  \beta \Phi_{b,\gamma}(F)=& -\ln{\sinh(\beta F b) \over \beta F b} + {\text{Li}_2(e^{-2\beta Fb}) - \xi(2) \over 2 \beta  \gamma  b}\\
                                                                   & - {F\over \gamma} [\ln (1-e^{-2\beta F b}) + {\beta F b \over 2} - 1],\
  \end{array} \\\\

 \begin{array}{ll}
  \beta \Phi_{{L^*}}(F)= & {x^2_{L^*}(F) \over 2A L^*}  - {x_{L^*}(F)+L^*\over 4A}\\ 
                                                          &+ {L^{*2} \over {4A(L^*-x_{L^*}(F))}}-{F x_{L^*}(F)\over k_\text{B}T}.\
 \end{array} \

\end{array}\right.
\end{equation}

Here, $\text{Li}_2(z)=\sum\limits_{k=1}^\infty {z^k \over k^2}$ is the second order polylogarithm function (also known as Jonquire's function), and $\xi(2) \sim 1.645$ is the Riemann-Zeta function evaluated at $z=2$. $k(F)$ is then determined by applying the Arrhenius law $k(F)  =k_\text{0} e^{-\beta \Delta \Phi^*(F)}$:
\begin{equation}
\label{Eq_k}
k(F)  =k_\text{0} e^{-\beta \left( \Phi_{b^*, \gamma^*}(F)-\Phi_{b^\text{0}, \gamma^\text{0}}(F)  +  \Phi_{{L^*}}(F)\right)}.
\end{equation}
At forces $\gg k_\text{B}T/b^\text{0}$, $\gg k_\text{B}T/b^*$ and $\gg k_\text{B}T/A$, $k(F)$ has a simple asymptotic expression:
\begin{equation}
\label{Eq_k1}
k(F)  = \tilde{k}_\text{0} e^{\beta \left( \sigma F+ \alpha F^2/2 - \eta F^{1/2}\right) },
\end{equation}
which contains a kinetics parameter $\tilde{k}_\text{0} = k_\text{0}{b^\text{0}\over b^*}e^{{\xi(2)\over 2}({k_\text{B}T\over \gamma^* b^*}-{k_\text{B}T\over \gamma^\text{0} b^\text{0}})}$, and three model parameters  $\sigma = L^*+(b^*-b^\text{0})-({k_\text{B}T\over \gamma^*} - {k_\text{B}T \over \gamma^\text{0}})$, $\alpha={b^*\over \gamma^*} - {b^\text{0}\over \gamma^\text{0}}$, and $\eta=L^* \sqrt{k_\text{B}T\over A}$. Typical values of ${k_\text{B}T\over \gamma^\text{0}}$ and ${k_\text{B}T\over \gamma^*}$ are in the range of $10^{-3}$ nm  - $10^{-2}$ nm (SI: S\rom{1}-\rom{2}, Tab. S1); therefore, $\sigma \sim L^*+(b^*-b^\text{0})$. An alternative derivation of Eq.~\ref{Eq_k1} is provided in Supporting Information (S\rom{4} : ``Alternative derivation of Eq.~\ref{Eq_k1}''). Here we emphasize that, since Eq.~\ref{Eq_k1} is an large-force asymptotic formula, $\tilde k_0$ should not be interpreted as the zero-force transition rate. The zero-force rate $k_0$ predicted by the model should be based on Eq.~\ref{Eq_k}, which is related to $\tilde{k}_\text{0}$ by the following equation:

\begin{equation}\label{Eq_k0}
k_\text{0}=\tilde{k}_\text{0}{b^* \over b^\text{0}}e^{-{\xi(2)\over 2}({k_\text{B}T\over \gamma^* b^*}-{k_\text{B}T\over \gamma^\text{0} b^\text{0}})}.
\end{equation}

For experiments that record transition force distribution $p(F)$ under a time-varying force constraint, it is straightforward to apply Eq.~\ref{Eq_k1} to fit such data by a simple transformation \cite{dudko2006}: 

\begin{equation}
p(F)=k(F)/\dot{F} \exp\left(-\int\limits_0^F \frac{k(F')}{\dot{F}} dF'\right),
\label{F_dist}
\end{equation}
where $\dot{F}$ is the time derivative of $F(t)$. $p(F)$ in Eq.~\ref{F_dist} is a density function, therefore the transition force histogram obtained from experiments should be reconstructed as $\langle$number of counts per bin$\rangle$ / $\langle$the total number of counts$\rangle$ / $\langle$bin size$\rangle$.

Clearly, in the three model parameters of Eq.~\ref{Eq_k1}, $\sigma$ is the contour length difference and $\alpha$ describes the deformability difference between the folded core of the transition state and the native state. $\eta$ only depends on the contour length of the flexible polymer in the transition state. Therefore, the best-fitting values of these parameters can provide important insights regarding how the different structural-elastic properties between the transition state and the native state affect $k(F)$. It is even possible to use this model to obtain further insights into the structural-elastic properties of the transition state based on the best-fitting values of $\sigma$, $\alpha$ and $\eta$. The native state structure is often known and therefore $b^\text{0}$ is determined. In addition, $\gamma^\text{0}$ can be estimated with reasonable accuracy using all-atom molecular dynamics (MD) simulations (SI: S\rom{1}-\rom{2}). Hence, for molecules with a known native state structure, the structural-elastic parameters of the transition state can be solved from $\sigma$, $\alpha$ and $\eta$.

\subsection{Applications in interpreting experimental data}
\begin{figure}[htbp]
\centering
\includegraphics[width=0.6\textwidth]{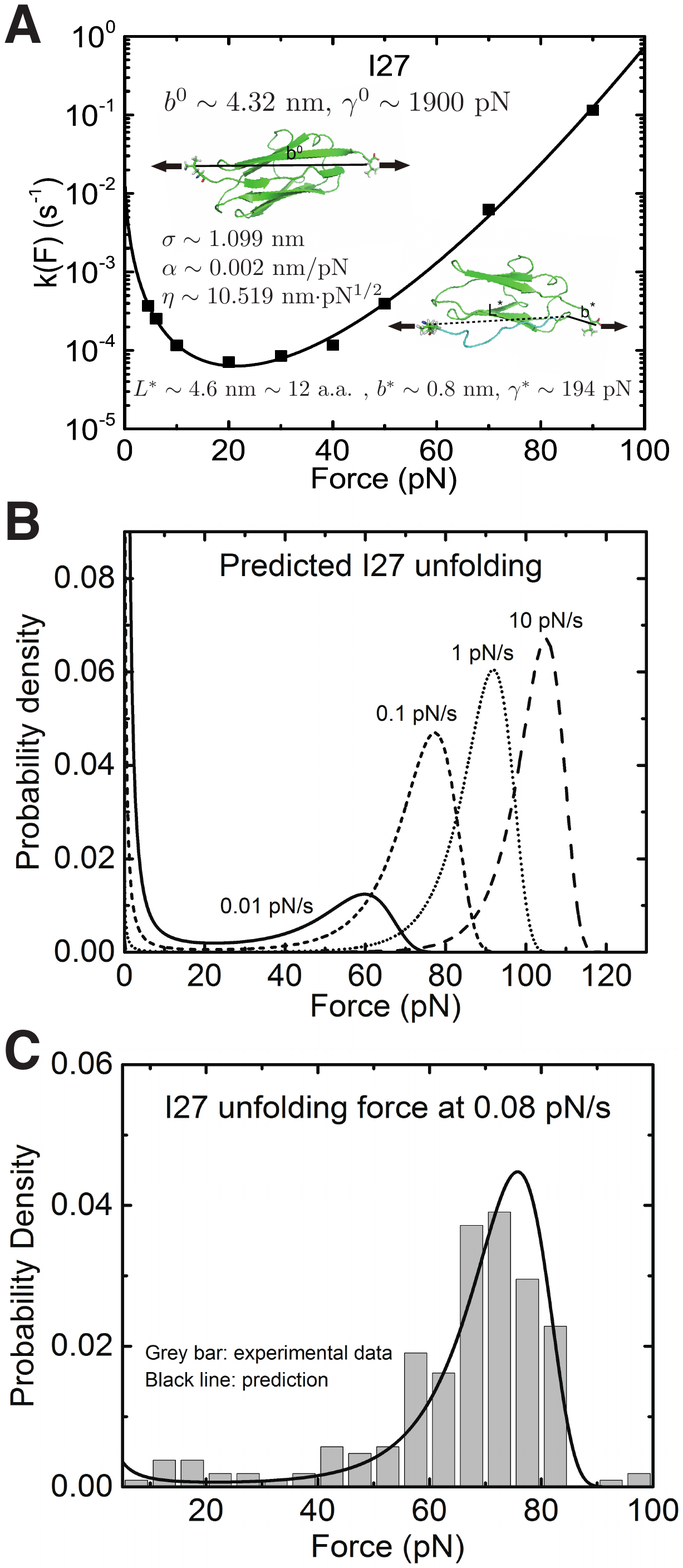}
\caption{{\bf Application of Eq.~\ref{Eq_k1} to interpret experimental data of titin I27.} (A) The $k(F)$ data for titin I27 domain unfolding \cite{Yuan-Chen2017} are indicated with black squares and fitted with Eq.~\ref{Eq_k1} (black line). The goodness-of-fit was evaluated by a R-Square of $\sim 0.997$ and a Root mean squared error (RMSE) of $\sim 0.162$. The best-fitting model parameters and the structural-elastic parameters determined based on the native state structure, steered MD simulation, or solved from the best-fitting parameters are indicated in the panel. (B) The panel shows the predicted I27 unfolding force distribution $p(F)$ using Eq.~\ref{F_dist} based on the best-fitting parameters for $k(F)$, with different loading rates of 0.01 pN/s (solid line), 0.1 pN/s (short dash line), 1 pN/s (short dot line) and 10 pN/s (dash line). (C) Comparison between the predicted $p(F)$ of I27  (solid black curve) and the experimental data (grey bars) shows good agreement at a loading rate of 0.08 pN/s.}
\label{Fig.I27}
\end{figure}

We firstly applied Eq.~\ref{Eq_k1} to fit $k(F)$ observed for titin I27 domain, and tested whether the fitting parameters can provide insights into how the structural-elastic properties of the molecule play a role in determining the transition kinetics. The titin I27 domain has a known transition state structure, which allows us to examine the quality of the prediction of the transition state properties based on the best-fitting parameters. As described earlier, the experimental data of I27 exhibits a ``catch-to-slip'' switching behaviour, where $k(F)$ switches from a decreasing function to an increasing function when force exceeds a certain threshold value at around 22 pN (Fig.~\ref{Fig.I27}A, black squares) \cite{Yuan-Chen2017}. At forces larger than $\sim$ 60 pN, the force-dependent unfolding rate converges to a Bell-like behaviour (Fig.~\ref{Fig.I27}A). The best-fitting parameters according to Eq.~\ref{Eq_k1} without any restriction are determined as: $\tilde k_0=0.026\pm 0.014$ s$^\text{-1}$, with 95\% confidence bounds of $(-0.020,0.073)$ s$^\text{-1}$; $\sigma = 1.099 \pm 0.243$ nm, with 95\% confidence bounds of $(0.510,1.689)$ nm; $\alpha = 0.002 \pm 0.003$ nm/pN, with 95\% confidence bounds of $(-0.004,0.007)$ nm/pN; and $\eta=10.519 \pm 1.542$ nm$\cdot$pN$^{1/2}$, with 95\% confidence bounds of $(6.356,14.682)$ nm$\cdot$pN$^{1/2}$. Here, the errors indicate standard deviations obtained with bootstrap analysis (SI: S\rom{5}, Tab. S2) and the 95\% confidence bounds are determined by fitting of all the data points (Fig.~\ref{Fig.I27}A, black squares). We also tested the robustness of the convergence of the fitting by repeating the fitting procedure with 10 different well-separated initial sets of values, and found that the best-fitting parameters converged to the same set regardless of the initial values (SI: S\rom{6}, Tab. S5). 

Based on the structure of I27 and steered MD simulations, $b^\text{0} \sim 4.32$ nm and $\gamma^\text{0} \sim 1900$ pN were estimated (SI: S\rom{1}-\rom{2}, Figs. S2 and S6). From the best-fitting parameters, $L^\text{*} = 4.6 \pm 0.7$ nm, $b^\text{*} = 0.8 \pm 0.4$ nm and $\gamma^\text{*} = 194 \pm 41$ pN were solved for the transition state. The value of $L^\text{*}$ corresponds to a peptide of $12 \pm 2$ residues, which is in good agreement with the previously known result that the transition state of I27 involves a peeled A-A$^\prime$ peptide chain of 13 residues (SI: Fig. S2) \cite{Yuan-Chen2017, Lu1998, Lu1999, Williams2003, Best2003}. This result shows that our model indeed can provide information of the structural-elastic properties of the transition state. The zero-force transition rate predicted by the model is estimated to be $k_\text{0} \sim 5\times 10^{-3}$ s$^{-1}$ according to Eq.~\ref{Eq_k0}. This value is consistent with that recently reported in \cite{Yuan-Chen2017} but differs from the value extrapolated based on Bell's model in earlier studies \cite{carrion1999mechanical} (see discussions in the discussion section).

Based on the best-fitting parameters, one can predict the I27 unfolding force probability density function $p(F)$ using Eq.~\ref{F_dist} at any loading rate. Figure~\ref{Fig.I27}B shows predicted $p(F)$ at several loading rates from 0.01 pN/s to 10 pN/s. We next compare the predicted $p(F)$ of I27 with experiments. Previous AFM experiments suggest that the native state of I27 transits to an intermediate state with the A strand detached from the B strand at forces $>$100 pN, and unfolding transition starts from this intermediate state at forces above 100 pN \cite{marszalek1999mechanical}. Since the $k(F)$ data in Fig.~\ref{Fig.I27}A were measured at forces below 100 pN, we chose to conduct experiment with a loading rate of 0.08 pN/s at which the unfolding forces are mainly below 100 pN for the comparison. Figure~\ref{Fig.I27}C shows the unfolding force density function constructed from 210 unfolding forces of I27 from 7 independent molecular tethers (vertical bars with a bin size of 5 pN) and the predicted $p(F)$ according to Eq.~\ref{F_dist} using the best-fitting values of the parameters ($\tilde k_0 = 0.026$ s$^\text{-1}$, $\sigma = 1.099$ nm, $\alpha = 0.002$ nm/pN and $\eta=10.519$ nm$\cdot$pN$^{1/2}$) described in the preceding section. The comparison shows good agreement between the predicted and experimental results.  

\begin{figure}[htbp]
\centering
\includegraphics[width=0.6\textwidth]{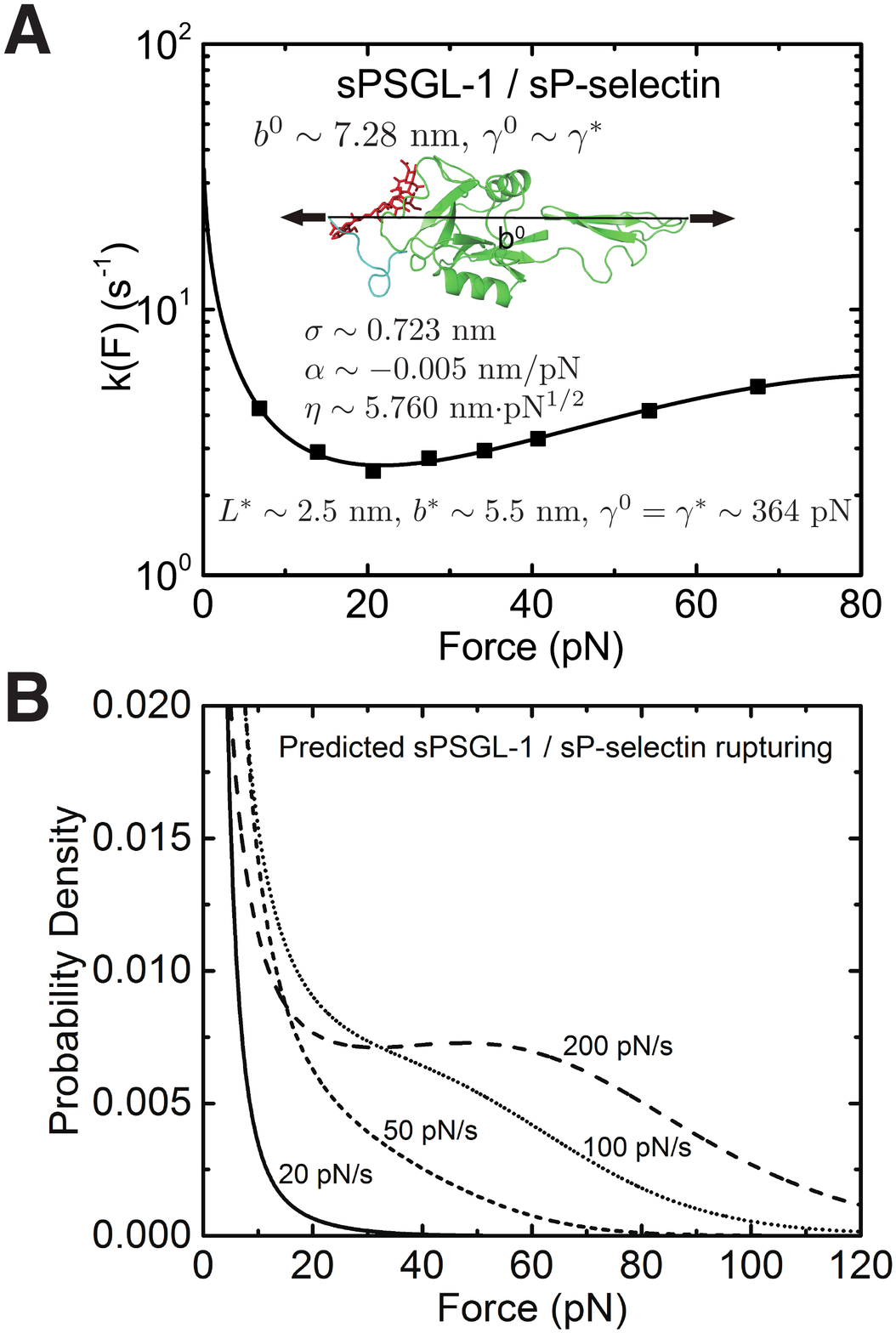}
\caption{{\bf Application of Eq.~\ref{Eq_k1} to interpret experimental data of monomeric PSGL-1/ P-selectin.} (A) The $k(F)$ data obtained for rupturing of monomeric PSGL-1/ P-selectin complex (Fig. $4b$ in Ref.~\cite{Marshall-Zhu2003}) are indicated with black squares and fitted with Eq.~\ref{Eq_k1} (black line). The goodness-of-fit was evaluated by a R-Square of $\sim 0.991$ and a Root mean squared error (RMSE) of $\sim 0.032$. The best-fitting model parameters and the structural-elastic parameters determined based on the native state structure, steered MD simulation, or solved from the best-fitting parameters are indicated in the panel. (B) The panel shows the predicted sPSGL-1 / sP-selectin  rupturing force distribution $p(F)$ using Eq.~\ref{F_dist} based on the best-fitting parameters for $k(F)$, with different loading rates of 20 pN/s (solid line), 50 pN/s (short dash line), 100 pN/s (short dot line) and 200 pN/s (dash line).} 
\label{Fig.PSGL}
\end{figure}

We next investigated the force-dependent rupturing rate of the monomeric PSGL-1/ P-selectin complex, which also demonstrates a ``catch-to-slip'' switching behaviour (Fig.~\ref{Fig.PSGL}A, black squares) \cite{Marshall-Zhu2003}. In addition, the $k(F)$ profile does not approach a Bell-like shape in the slip bond region when force is further increased. Therefore, this protein complex represents a more complicated situation compared with I27. The best-fitting parameters without any restriction are determined as  $\tilde k_0=51.786 \pm 27.083$ s$^\text{-1}$, with 95\% confidence bounds of $(12.499, 91.072)$ s$^\text{-1}$;  $\sigma=0.723 \pm 0.162$ nm, with 95\% confidence bounds of $(0.468, 0.978)$ nm;  $\alpha=-0.005 \pm 0.001$ nm/pN, with 95\% confidence bounds of $(-0.008, -0.002)$ nm/pN; and $\eta=5.760 \pm 1.275$ nm$\cdot$pN$^{1/2}$, with 95\% confidence bounds of $(4.019, 7.501)$ nm$\cdot$pN$^{1/2}$. The errors and the robustness of the parameter convergence are generated/tested similar to the case of I27 (SI: S\rom{5}-\rom{6}, Tabs. S3 and S6). 

$b^\text{0} \sim 7.28$ nm was determined based on the structure of the PSGL-1/ P-selectin complex (SI: Fig. S3). As P-selectin occupies most of the volume of the complex, its stiffness should be the determining factor for the deformability of the folded structure/core for both the native state and the transition state (i.e., $\gamma^\text{0} \sim \gamma^*$). From these values, $L^* = 2.5 \pm 0.6$ nm, $b^* = 5.5 \pm 0.4$ nm, and $\gamma^\text{0} = \gamma^* = 364 \pm 48$ pN were solved. These results predict a partially peeled peptide/sugar polymer in the transition state, which suggests that detachment of the sugar molecule covalently linked to the PSGL-1 from P-selectin is a necessary step that has to take place before rupturing (SI: Fig. S3). The zero-force transition rate predicted by the model is estimated to be $k_\text{0} \sim 39.1$ s$^{-1}$ according to Eq.~\ref{Eq_k0}. 

The predicted $p(F)$ using Eq.~\ref{F_dist} at several loading rates from 20 pN/s to 200 pN/s are shown in Fig.~\ref{Fig.PSGL}B. To the best of our knowledge, loading rate-dependent $p(F)$ for the rupturing of monomeric PSGL-1/P-selectin complex has not been experimentally measured in the force range similar to the $k(F)$ data; therefore, the predicted $p(F)$ in Fig.~\ref{Fig.PSGL}B will be awaiting for future experimental tests.

\begin{figure}[htbp]
\centering
\includegraphics[width=0.6\textwidth]{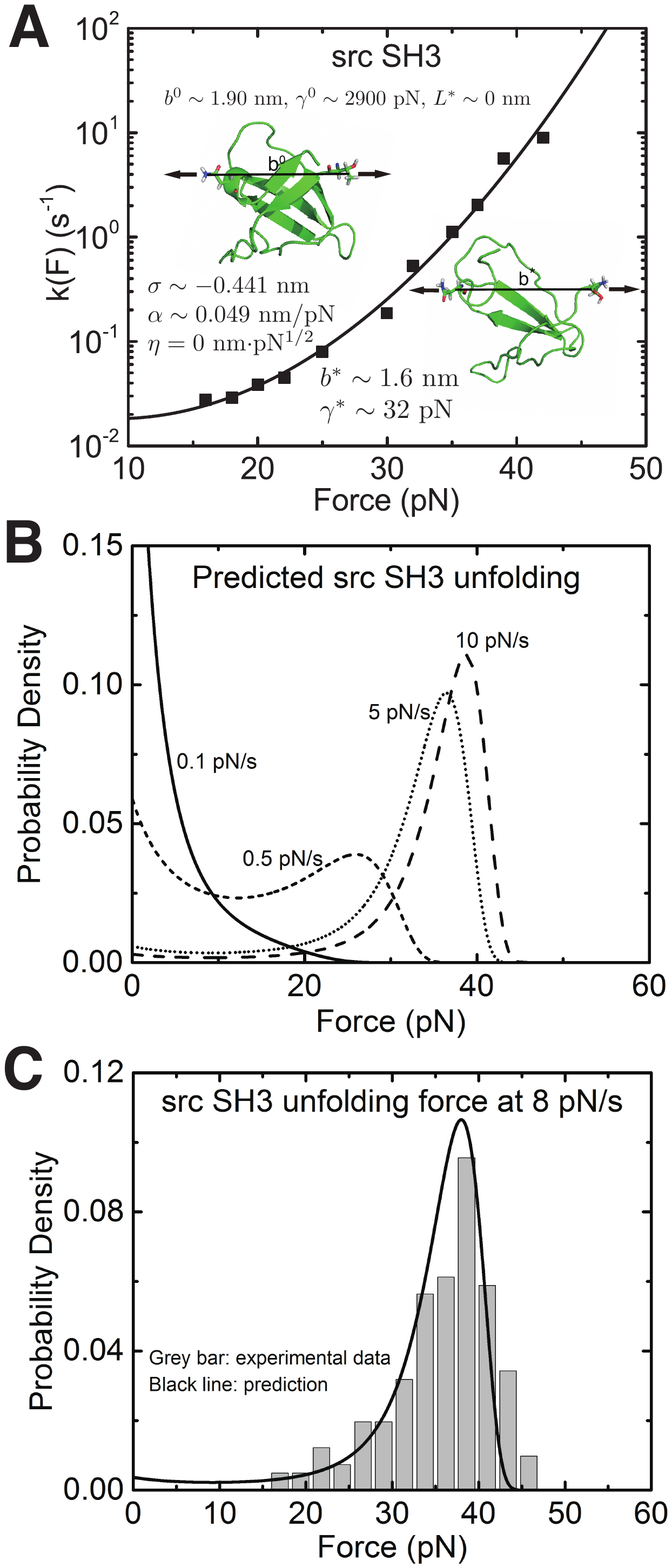}
\caption{{\bf Application of Eq.~\ref{Eq_k1} to interpret experimental data of src SH3.} (A) The $k(F)$ data obtained for src SH3 (Fig. $3A$ in Ref.~\cite{Jagannathan2012}) are indicated with black squares and fitted with Eq.~\ref{Eq_k1} (black line). The goodness-of-fit was evaluated by a R-Square of $\sim 0.992$ and a Root mean squared error (RMSE) of $\sim 0.224$.  The best-fitting model parameters and the structural-elastic parameters determined based on the native state structure, steered MD simulation, or solved from the best-fitting parameters are indicated in the panel. (B) This panel shows the predicted src SH3 unfolding force density function $p(F)$ using Eq.~\ref{F_dist} based on the best-fitting parameters for $k(F)$, at different loading rates of 0.1 pN/s (solid line), 0.5 pN/s (short dash line), 5 pN/s (short dot line) and 10 pN/s (dash line). (C) The predicted $p(F)$ of src SH3 (solid black curve) agrees with the previously published experimental data (Fig. $2B$ in Ref.~\cite{Jagannathan2012}) (grey bars) at a loading rate of 8 pN/s.} 
\label{Fig.SH3}
\end{figure}

We also applied the theory to understand the unfolding of the src SH3 domain under a special stretching geometry that causes a significant deviation from Bell's model (Fig.~\ref{Fig.SH3}A, black squares) \cite{Jagannathan2012} On the logarithm scale, it exhibits a convex profile increasing with force, which strongly suggests that the $\alpha F^2 \over 2$ term in the exponential of Eq.~\ref{Eq_k1} with a positive $\alpha$ is the cause of the observed $k(F)$. Unconstrained fitting resulted in a negative value of $b^*$, which is physically impossible. We found that $\eta < 4.3$ is needed to ensure a positive $b^*$. Good quality of fitting was obtained for any values of $\eta < 4.3$ (SI: S\rom{7}), suggesting that the length of peptide produced in the transition state is not responsible for the observed $k(F)$ data. The value of  $\alpha \sim 0.042 - 0.048$ nm/pN is insensitive to changes in $\eta$ (SI: Tab. S8), strongly suggesting the deformability of the folded core in the transition state as the key factor of the observed $k(F)$ .  

In order to further obtain more accurate structural-elastic properties of the transition state of src SH3 domain, additional information of the peptide length in the transition state is needed. Previous study estimated a small transition distance $\sim 0.45$ nm in the force range of 15-25 pN \cite{Jagannathan2012}, suggesting insignificant fraction of peptide in the transition state (SI: S\rom{7}). Consistently, our steered MD simulation shows a negligible production of peptide under force during transition (SI: Fig. S4). Based on these information, we estimated $b^* $ and $\gamma^*$ by approximating $\eta \sim 0$. The resulting best-fitting parameters are determined as $\tilde{k}_\text{0} = 0.030 \pm 0.043$ s$^\text{-1}$, with 95\% confidence bounds of $(-0.033, 0.092)$ s$^\text{-1}$; $\sigma = -0.441 \pm 0.249$ nm, with 95\% confidence bounds of $(-1.083, 0.202)$ nm; and  $\alpha = 0.049 \pm 0.009$ nm/pN, with 95\% confidence bounds of $(0.027, 0.071)$ nm/pN. The errors and the robustness of the parameter convergence are generated/tested similar to the case of I27 (SI: S\rom{5}-\rom{6}, Tabs. S4 and S7). 

The structural-elastic parameters of the native state were determined to be $b^\text{0} \sim 1.90$ nm and $\gamma^\text{0} \sim 2900$ pN based on the structure and steered MD simulations (SI: S\rom{1}-\rom{2}, Figs. S4 and S7). Finally, based on the best-fitting values, $b^* = 1.6 \pm 0.2$ nm and $\gamma^* = 32 \pm 9$ pN were solved. The estimated value of $\gamma^*$ is reasonably in agreement with the value estimated based on steered MD simulations for the transition state of src SH3 (SI: Fig. S7).

The predicted $p(F)$ for src SH3 using Eq.~\ref{F_dist} at several loading rates from 0.1 pN/s to 10 pN/s are shown in Fig.~\ref{Fig.SH3}B. The unfolding force histogram of SH3 was measured at a loading rate of 8 pN/s \cite{Jagannathan2012}, which was converted to probability density function. The comparison between the experimental data and $p(F)$ predicted by Eq.~\ref{F_dist} using the best-fitting parameters reported in this study shows very good agreement (Fig.~\ref{Fig.SH3}C). 

\begin{figure}[htbp]
\centering
\includegraphics[width=0.6\textwidth]{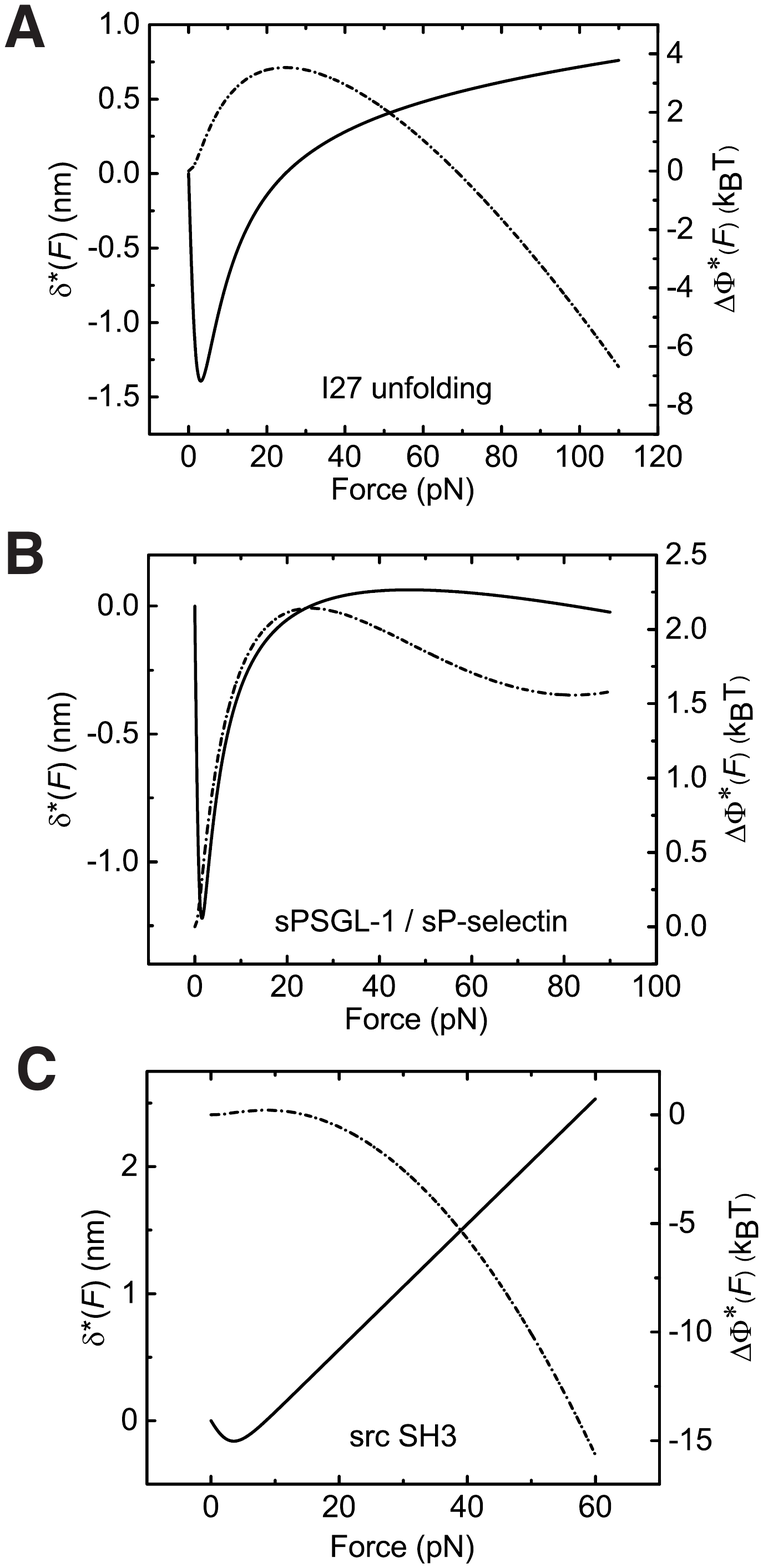}
\caption{{\bf{Force-dependent transition distance and change of free energy barrier.}} The force-dependent transition distance $\delta^\text{*}(F)$ (solid line) calculated by Eq.~\ref{Eq_td} and the force-dependent change of the free energy barrier $\Delta \Phi^\text{*}(F)$ (dash dot line) calculated by Eq.~\ref{Eq_DPhi} for I27 (A), monomeric PSGL-1/ P-selectin (B) and src SH3 (C) are shown. $\delta^\text{*}(F)$ and $\Delta \Phi^\text{*}(F)$ are calculated based on the values of the five structural-elastic parameters ($b^\text{0}$, $\gamma^\text{0}$, $b^\text{*}$, $\gamma^\text{*}$, $L^\text{*}$) determined based on the best-fitting parameters ($\sigma$, $\alpha$, $\eta$), the molecular structures and steered MD simulations for the respective molecules described in the Results section.} 
\label{Fig.dx}
\end{figure}

As shown in the previous paragraphs, the five structural-elastic parameters ($b^\text{0}$, $\gamma^\text{0}$, $b^\text{*}$, $\gamma^\text{*}$, $L^\text{*}$) for I27, monomeric PSGL-1/ P-selectin and src SH3 are determined based on the best-fitting model parameters ($\sigma$, $\alpha$, $\eta$), the molecular structures and steered MD simulations. With these structural-elastic parameters, the force-dependent transition distance $\delta^\text{*}(F)$ and the change of the free energy barrier $\Delta \Phi^\text{*}(F)$ can be computed using Eq.~\ref{Eq_td} and Eq.~\ref{Eq_DPhi} (Fig.~\ref{Fig.dx}). The results reveal that the three molecules have markedly different profiles of $\delta^\text{*}(F)$ and $\Delta \Phi^\text{*}(F)$.  For all the three molecules, the complex shapes of $\delta^\text{*}(F)$ over 1-100 pN force range deviate from Bell's model that assumes a force-independent transition distance. These complex profiles of $\delta^\text{*}(F)$ result in complex force-dependent changes of free energy barrier ($\Delta \Phi^\text{*}(F)$), which in turn affects the force-dependence of the transition rate in a very complex manner. For I27 and PSGL-1/ P-selectin, the transition distances can become negative over a broad force range up to $\sim 20$ pN, which results in a ``catch-bond'' behaviour at forces below 20 pN. Remarkably, the force-dependent transition distance drops dramatically when force increases from 0 pN to a few pN. These behaviours of the force-dependent transition distance are a result from the highly flexible nature of the peptide chain with a contour length $L^*$ and a persistence length $A\sim 0.8$ nm \cite{Winardhi2016} produced in the transition state. According to the WLC polymer model \cite{Marko1995}, the force-extension curve of a peptide polymer at low force regime ($F< k_\text{B}T/A \sim 5$ pN) can be approximated by a Hookean spring with a spring constant of ${3\over 2}{k_\text{B}T \over AL^*}$, which are $\sim 1.7$ pN/nm for I27 and $\sim 3.1$ pN/nm for PSGL-1/ P-selectin based on the respective best-fitting values of $L^*$. 

\section{Discussion}

In summary, we have derived a novel analytical expression of $k(F)$ for single-barrier transitions. Most importantly, the parameters are functions of the structural-elastic parameters of the molecules; therefore, their values directly inform us about the structural-elastic properties of the molecule. We have shown that it is possible to determine the structural-elastic parameters of the molecule in both the native and the transition states by combining this model with the steered MD simulations. 

In our previous publication \cite{Yuan-Chen2017}, based on the prior knowledge of the crystal structure of the native state of I27 (PDB ID:1TIT) and the structure of its transition state suggested from MD simulations \cite{Lu1998,Lu1999,Best2003}, we calculated $\Delta \Phi^\text{*}(F)$ without any model parameters, with an assumption that both the native state and the folded portion of the transition state are non-deformable. Applying the Arrhenius law, we showed that this parameter-free $\Delta \Phi^\text{*}(F)$ correctly describes the shape of the experimentally measured $k(F)$ up to 100 pN. The only free parameter in that calculation is the attempting rate $k_\text{0}$, which only affects the value of $k(F)$ (i.e., this parameter is unrelated to the force-dependence of $k(F)$). The work described in this paper differs from that earlier work in that: 1) it does not require prior knowledge of the structural-elastic properties of the molecule, 2) the best-fitting parameters ($\sigma$, $\alpha$ and $\eta$) reflect differences in the structural-elastic properties of the molecule between the transition and native states, and 3) with additional knowledge on the structural-elastic properties of the native state that can often be obtained from crystal structure and MD simulations, these best-fitting parameters can provide important information about the nature of the transition state. 

In most of experiments, $k(F)$ is measured over certain force range. Fitting to the data based on any kinetics model, it is attempting to extrapolate the fitted $k(F)$ to forces beyond the experimentally measured range. However, this is dangerous if the force extrapolated to is far away from the experimentally measured range. This is because the nature of the transition may vary with the force, while most of the models \cite{Evans1997, hummer2003kinetics, dudko2006, cossio2016kinetic} , including ours, are derived based on assuming a unique initial folded state and a single transition barrier. Such assumption may only be valid in limited force range. For example, previous AFM experiments and MD simulations \cite{marszalek1999mechanical, Lu1999} suggest that at forces below 100 pN, the initial folded state of I27 has all the seven $\beta$-strands folded in the native structure. However, at forces $>100$ pN, the initial folded state transits to an intermediate state with the A strand detached from the B strand \cite{marszalek1999mechanical, Lu1999}. Therefore, $k(F)$ fitted based on experimental data at forces below 100 pN should not be extrapolated to forces above 100 pN and vice versa. As an example, the zero-force transition rate of I27 was estimated to be $\sim 0.0005$ s$^{-1}$ by extrapolating experimental data obtained at forces above 100 pN according to Bell's model \cite{carrion1999mechanical}, which is about 10 times slower than that estimated by our model. However, that extrapolation did not take into consideration of the difference in the initial states between forces below and above 100 pN. In addition, Bell's model cannot describe the recently reported ``catch-bond'' behaviour of I27 at forces below 22 pN \cite{Yuan-Chen2017}, which further contributes to the discrepancy. 

The simple expression of Eq.~\ref{Eq_k1} is derived based on large force asymptotic expansion ($F \gg k_\text{B}T/b^\text{0}$, $F\gg k_\text{B}T/b^*$ and $F\gg k_\text{B}T/A$). The typical sizes of protein domain and the folded core in the transition state are in the order of a few nanometers; therefore, $k_\text{B}T/b^\text{0}$ and  $k_\text{B}T/b^*$ are close to 1 pN. If in the transition state a protein peptide or a ssDNA/ssRNA polymer is produced, due to their very small bending persistence of $A \sim 1$ nm \cite{Winardhi2016, bosco2013elastic}, $k_\text{B}T/A\sim 5$  pN becomes the predominating factor that imposes a restriction to the lower boundary of force range to apply Eq.~\ref{Eq_k1}. In actual application, the applicable forces do not have to be much greater than 5 pN, since the force-extension curve of a flexible polymer with $A\sim 1$ nm calculated based on the asymptotic large force expansion differs from the one according to the full Marko-Siggia formula \cite{Marko1995} by less than 10\% at forces above 3 pN (SI: Fig. S9). Therefore, Eq.~\ref{Eq_k1} can be applied to forces $> 3$ pN. Consistently, we have shown that Eq.~\ref{Eq_k1} can fit three different experimental data in this force range.

Since Eq.~\ref{Eq_k1} is not applicable at forces $\ll 3$ pN, $\tilde{k}_\text{0}$ should not be interpreted as the transition rate at zero force. Under cases where the five physical parameters ($b^\text{0}$, $\gamma^\text{0}$, $b^\text{*}$, $\gamma^\text{*}$ and $L^\text{*}$) can be solved from the best-fitting parameters ($\sigma$, $\alpha$ and $\eta$), extrapolation to lower forces is possible using the complete solution of Eq.~\ref{Eq_k}. A better quantity that is more indicative of zero force transition rate is $k_\text{0}$ in Eq.~\ref{Eq_k} that is derived without using asymptotic large force expansion, which can be computed based on the best fitting value of $\tilde{k}_\text{0}$ according to Eq~\ref{Eq_k0}. However, caution should still be taken for such extrapolation since at very low forces the WLC model of the flexible protein peptide or ssDNA/ssRNA may no longer be valid due to potential formation of secondary structures on these polymers.

The effects of the elastic properties of molecules on the force-dependent transition rate have been discussed in several previous works \cite{dembo1988reaction, valle2015elastic}. In particular, in a pioneering work published by Dembo et al.\cite{dembo1988reaction}, by treating the native and the transition states as molecular springs with different mechanical stiffness and lengths, the authors were able to predict the existence of catch, slip and ideal bonds. However, that model is too simple to explain complex $k(F)$ such as the ``catch-to-slip'' behaviour. In addition, treating the native and the transition states as molecular springs makes it impossible to relate the force dependence of transition rate to the actual structural parameters of the molecules in the native and transition states. For instance, it cannot predict whether there is a peptide produced in the transition state. In another work by Cossio et al. \cite{cossio2016kinetic}, the authors discussed a free energy landscape that has a force-dependent transition distance, based on which $k(F)$ was derived by applying the Kramers kinetics theory. A phenomenological form of the force-dependent transition distance is proposed to describe the kinetic ductility that results in a monotonically decreased transition distance as a function of force, which could only describe transition with ``slip'' kinetics. Different from these previous studies, our derivation is based on the structural-elastic properties of molecules in the transition state and the native state. Therefore,  its force dependence can be much richer. Depending on the structural-elastic properties of the molecules, the resulting force-dependent transition distance can be an increasing, decreasing or non-monotonic function of force. 

The analytical expressions of $k(F)$ (Eq.~\ref{Eq_k} and Eq.~\ref{Eq_k1}) are derived by applying Arrhenius law based on the structural-elastic parameters of molecules. The resulting relation between the rate and the force-dependent transition distance, $k(F)=k_\text{0}e^{\beta \int\limits_0^F \delta^*(F')dF'}$, is identical to that obtained in the framework of the Kramers theory \cite{dudko2008theory}. However, they differ from each other in a key aspect: In our theory $\delta^*(F)$ is calculated based on the structural-elastic parameters of molecules; therefore it does not involve describing the system using any transition coordinate and it does not depend on the dimensionality of the system. In contrast, in the framework of Kramers theory, $\delta^*(F)$ has to be calculated based on a presumed one-dimentional free energy landscape that must be expressed by the extension as the transition coordinate. As a result, $\delta^*(F)$ depends on the structural-elastic parameters of the molecules in our theory, while it relies on the parameters associated with shapes of the presumed one-dimension free energy landscape in the framework of the Kramers theory \cite{dudko2008theory}. Owing to this difference, our theory can be applied to a broader scope of experimental cases and the best-fitting parameters can provide important insights into the structural-elastic properties of the molecules in the native and the transition states. The three molecules selected to test the application of $k(F)$ derived in this work have markedly different profiles. The fact that the expression of $k(F)$ is able to perfectly fit the experimental data for all the three molecules reveals an exquisite interplay between the structural-elastic properties of molecules and the force-dependent transition rate.

\section{Methods}
\textit{Titin I27 domain unfolding experiments} -- A vertical magnetic tweezers setup \cite{Chen2011a} was used for conducting in vitro titin I27 domain stretching experiments. The sample protein (8I27) was designed with eight repeats of titin I27 domains spaced with flexible linkers (GGGSG) between each domain; The 8I27 was labeled with biotin-avi-tag at the N-terminus and spy-tag at the C-terminus. The expression plasmid for the sample protein was synthesised by geneArt. In a flow channel, the C-terminus of the protein was attached to the spycatcher-coated bottom surface through specific spy-spycather interaction, while the N-terminus was attached to a streptavidin-coated paramagnetic bead (2.8 $\mu$m in diameter, Dynabeads M-270) through specific biotin-streptavidin interaction. During experiments, the force on a single protein tether was linearly increased from $\sim 1$ pN up to $\sim 120$ pN with a loading rate of $\sim 0.08$ pN/s, to allow the unfolding of each I27 domain; after unfolding of the domains, the force was decreased to $\sim 1$ pN for $\sim 60$ sec to allow refolding of the domains before next force-increase scans. Each I27 unfolding events and its corresponding unfolding force were detected by a home-written step-finding algorithm. All experiments were performed in buffered solution containing 1$\times$ PBS, 1\% BSA, 1 mM DTT, at 22 $\pm$ 1 $^{\circ}$C. Additional information of the magnetic tweezers setup, force calibration, step-finding algorithm, protein sequences, protein expression, and flow channel preparation can be found in previous publications \cite{Chen2011a, Chen2015, Yuan-Chen2017}.

\textit{MD simulations} -- The all-atom molecular dynamics (MD) simulations used to estimate the value of $\gamma$ of the folded structure are introduced in the Supporting Information (SI: S\rom{1}-\rom{2}).

\textit{Data extraction} -- The data of $k(F)$ for monomeric PSGL-1/P-selectin complex and src SH3 domain, and the histogram of unfolding force for src SH3 were obtained by digitizing previously published experimental data ( Fig. $4b$ in \cite{Marshall-Zhu2003} for PSGL-1/P-selectin data, and Fig. $3A$ and Fig. $2B$ in \cite{Jagannathan2012} for src SH3 data). The values of $k(F)$ and the histogram of unfolding force were extracted using ImageJ with the Figure Calibration plugin developed by Frederic V. Hessman from Institut f\"ur Astrophysik G\"ottingen. 
 
\section{Author Contributions}
J.Y., S.G. and H.C. developed the theory. Q.T. performed steered MD simulations. S.G. and M.Y. performed the calculation and data fitting. Q.T. and S.L. performed the titin I27 domain unfolding experiments. J.Y. and S.G. wrote the paper. J.Y. conceived and supervised the study. 

\section{Acknowledgement}
The authors thank Jacques Prost (Institut Curie) for many stimulating discussions. This work was supported by the National Research Foundation (NRF), Prime Minister's Office, Singapore under its NRF Investigatorship Programme (NRF Investigatorship Award No. NRF-NRFI2016-03 to JY) and through the Mechanobiology Institute (to JY), Human Frontier Science Program (RGP00001/2016 to JY), and the National Nature Science Foundation of China (11474237 to HC).

\bibliography{Kinetics}




\onecolumngrid

\newpage

\renewcommand{\theequation}{S\arabic{equation}}
\renewcommand{\thefigure}{S\arabic{figure}}
\renewcommand{\thetable}{S\arabic{table}}
\setcounter{figure}{0}

\appendix*
\section{Supplemental Information}

The supporting information describes: 1) the steered molecular dynamics simulation method (S\rom{1}); 2)the extraction of the structural and elastic parameters of the native and the transition states of molecules investigated in this paper (S\rom{2});  3) the conformational free energy of a molecule under external force (S\rom{3}); 4) an alternative derivation of Eq. 5 (S\rom{4});  5) bootstrap analysis to determine fitting errors (S\rom{5}); 6) the robustness of convergence of the best-fitting parameters (S\rom{6}); and 7) fitting of Eq. 5 to experimental data of src SH3 domain over different presupposed peptide length in the transition state (S\rom{7}).

\subsection*{S\rom{1}. Molecular dynamics simulations}
All-atom molecular dynamics (MD) simulations were performed in Gromacs 5.1.1 \cite{abraham2015gromacs} with Parmbsc1 force field \cite{ivani2016parmbsc1} for DNA and with CHARMM36 force field \cite{best2012optimization} for proteins. Molecular structures of DNA is built by x3DNA software \cite{lu20033dna}, and the structures of titin I27 domain (PDB: 1tit) \cite{improta1996immunoglobulin}, src SH3 domain (PDB: 1srl) \cite{yu19931h}, monomeric PSGL-1/P-selectin complex (PDB: 1g1s) \cite{somers2000insights} are public data from protein data bank. All the simulations used explicit water TIP3P \cite{jorgensen1983comparison} with 150 mM NaCl to mimic physiological condition. Simulation boxes were heated to 300 K and then kept at constant temperature and pressure for 200 ps to relax. During steered molecular dynamics simulations, a constant force was applied to the force-bearing residues, therefore the end-to-end distance (extension) fluctuation of the molecules could be analysed. Standard deviation and mean value of extension were calculated from the last 20 ns of simulation. 

\begin{figure}[htbp]
\centering
\includegraphics[width=0.4\textwidth]{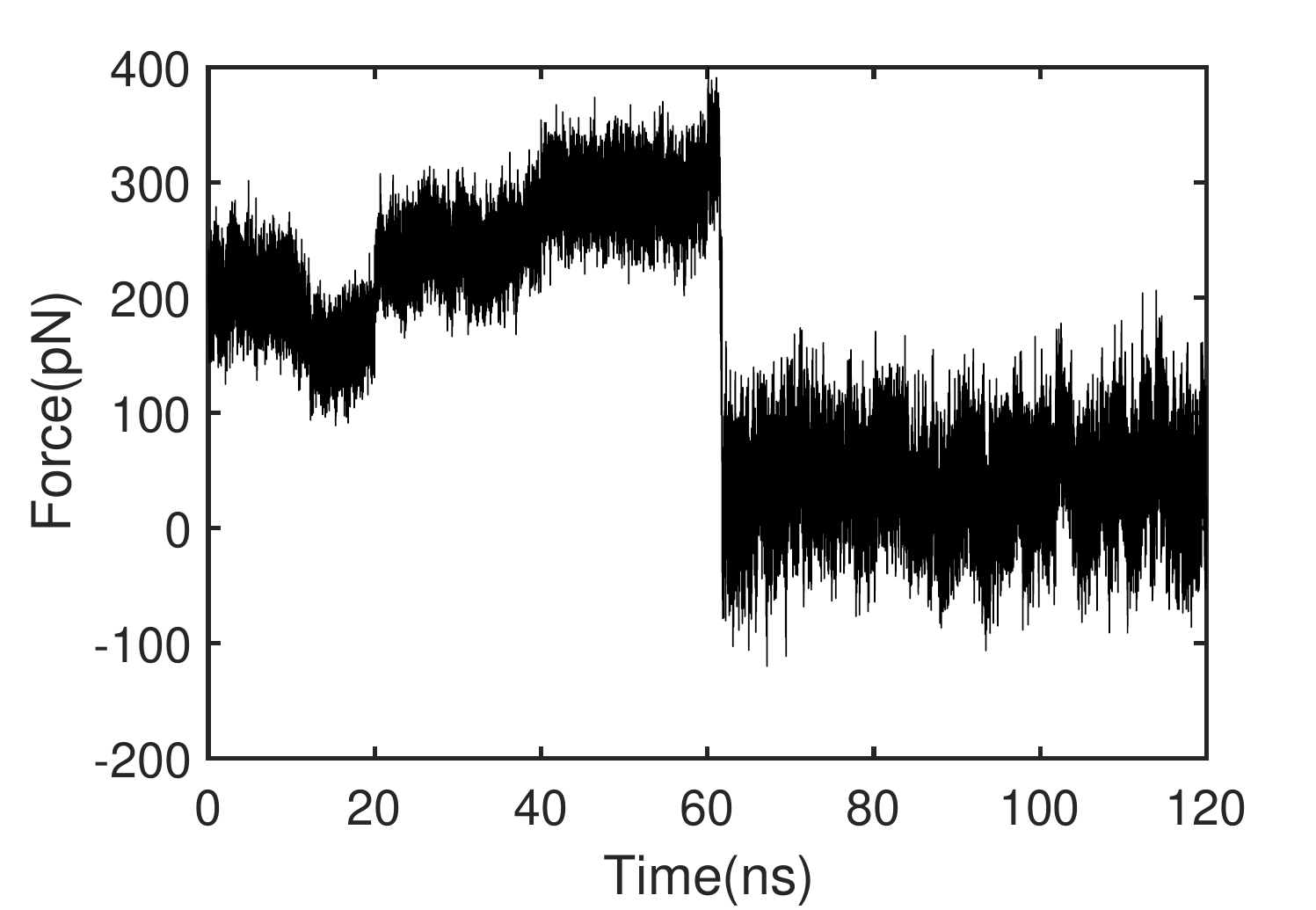}
\caption{{\bf Force applied on src SH3 domain during steered MD simulation.} Forces from a sequence of simulation (20 ns each) with increasing harmonic trap separation from 2.1nm-2.6nm were concatenated. At beginning, stepwised increase in force was observed as trap separation increased. As the trap separation increased to 2.4nm, the force firstly increased and suddenly dropped off, indicating a structural transition in the protein domain. The force drop was followed by a much weaker dependence of force on trap separation as the separation continued to increase, indicating a very different structure produced under stretching. Thus the structure after force drop was characterised as a transition state. }
\label{fig:sh3_force}
\end{figure}

The transition state of src SH3 is determined by steered MD simulations. A sequence of harmonic traps with same stiffness of 1000 pN/nm and different center separation of 2.1-2.6 nm were applied to the same force-bearing residues as in experiment \cite{Jagannathan2012}. With this stretching setup, the force slowly build up between the stretching residues, and the structure has enough time to relax to equilibrium. The force on stretching residues were recorded and concatenated (Fig.~\ref{fig:sh3_force}). Structural transition is indicated by the force drop occured at a trap separation of 2.4 nm. The structure after force drop was regarded as a transition state of the protein domain during unfolding.

\subsection*{S\rom{2}. Structural and elastic properties of the native and the transition states of molecules}
\begin{figure}[htbp]
\centering
\includegraphics[width=0.4\textwidth]{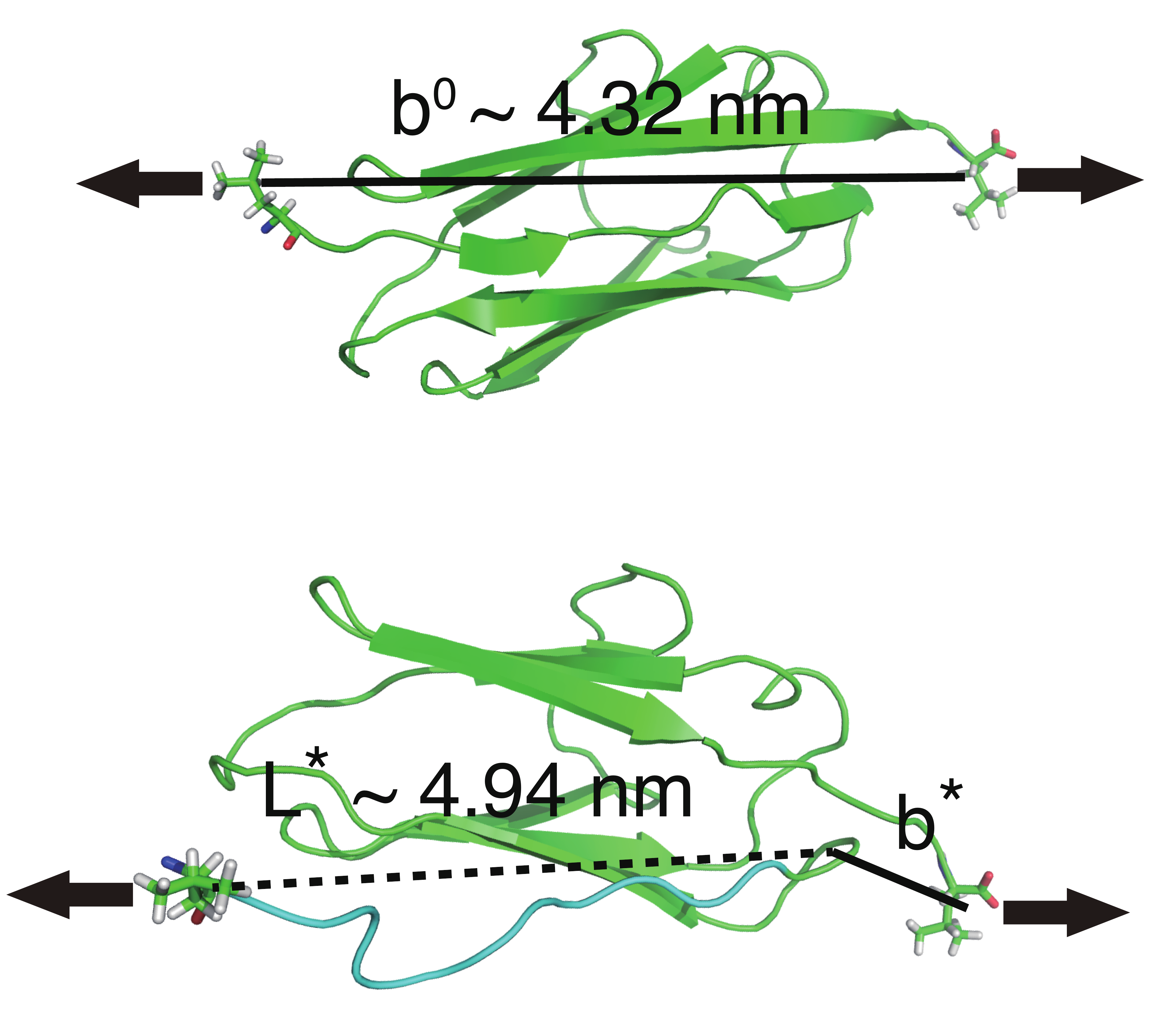}
\caption{{\bf The structure of titin I27 domain in native state and transition state.}  (A) The native state of I27 is a folded structure with $b^\text{0}=4.32$ nm. (B) The transition state of I27 is composed of a peptide of 13 residues \cite{Lu1998,Lu1999,Best2003,Williams2003} under force and a folded core with a relaxed length of $b^*=0.64$ nm.}
\label{fig:I27_rigid_body}
\end{figure}

\begin{figure}[htbp]
\centering
\includegraphics[width=0.4\textwidth]{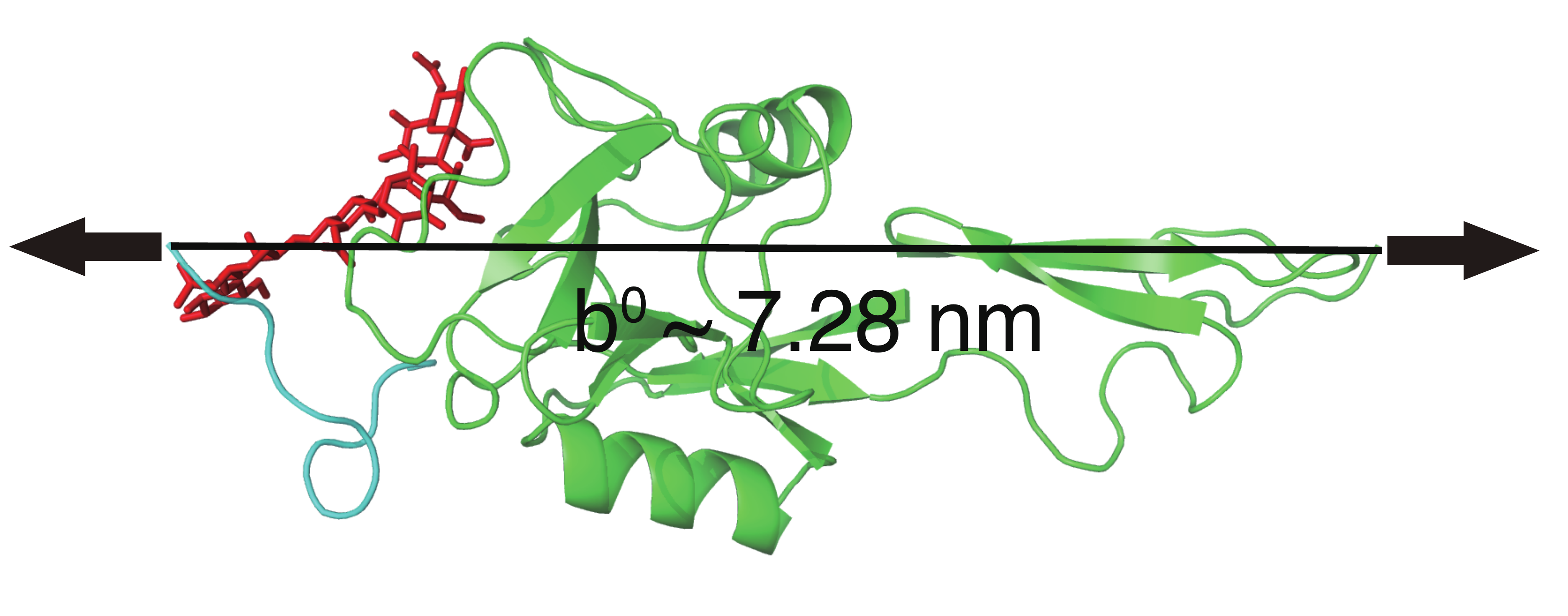}
\caption{{\bf The structure of monomeric PSGL-1/ P-selectin complex in native state.}  The protein complex has a folded structure length of $b^\text{0}=7.28$ nm. It contains a SLe$^\text{x}$ sugar chain (red) covalently linked to PSGL-1 (cyan) that binds to P-selectin (green).}
\label{fig:selectin_b}
\end{figure}

\begin{figure}[htbp]
\centering
\includegraphics[width=0.4\textwidth]{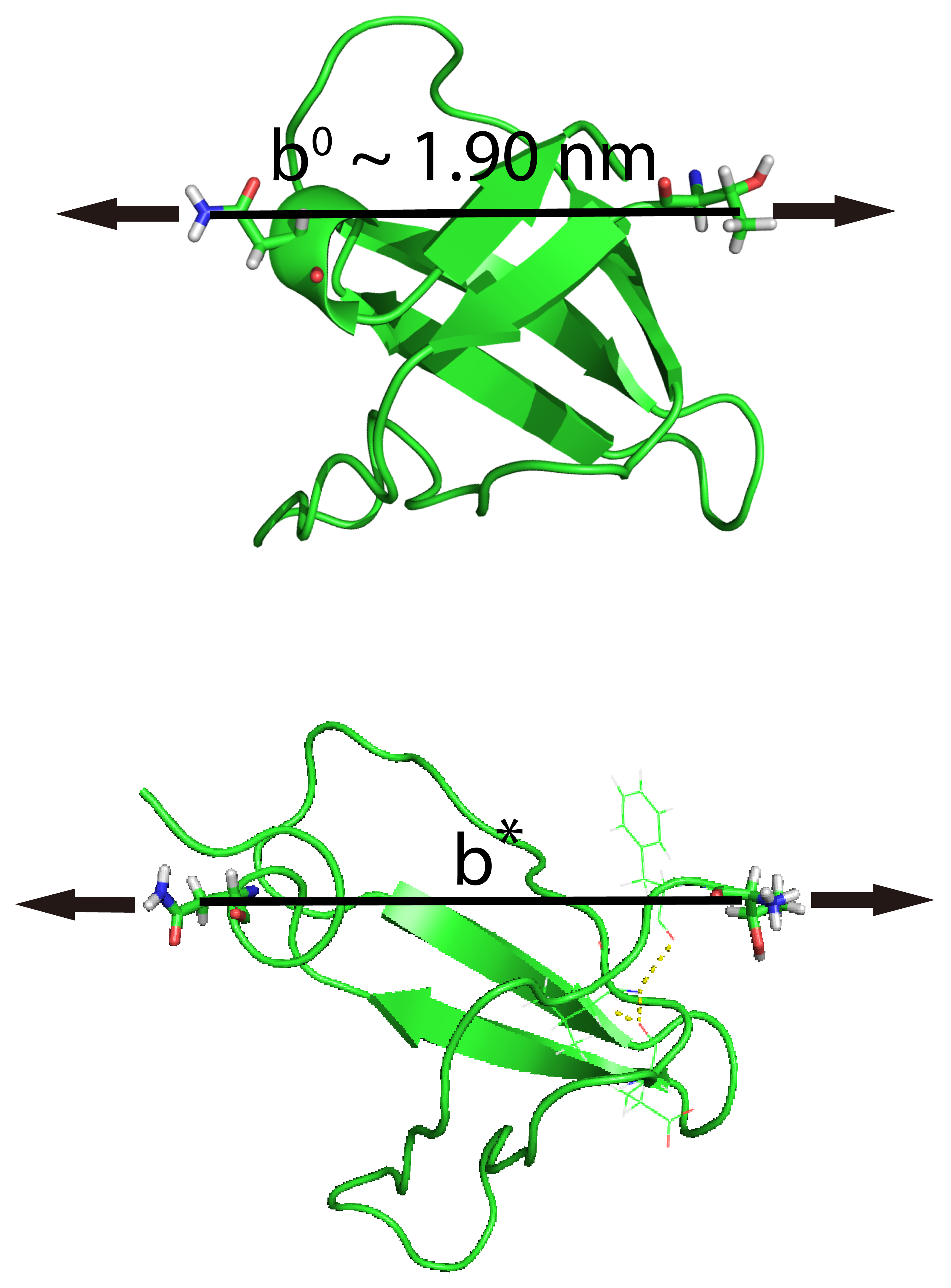}
\caption{{\bf The structure of src SH3 in native state and transition state under force.} (A) In the native state, the distance between force-bearing residues is $1.90$ nm, thus it is regarded as a folded structure with a relaxed length of $b^\text{0}=1.90$ nm. (B) A snapshot of the transition state produced by sequential stretching by harmonic traps (SI: Sec \rom{1}). The hydrogen bonds in N-terminal remains(as shown in yellow dashed lines, key residues involved are shown in line representation), while the C-terminal peptide peels off, which is not under force, so the released peptide under force is negligible.}
\label{fig:sh3_rigid_body_n_transition}
\end{figure}

The contour length of the folded structure in native state or the folded core in transition state of the molecules were estimated based on structures of the molecules (Fig.~\ref{fig:I27_rigid_body}-\ref{fig:sh3_rigid_body_n_transition}). Molecular dynamics simulations were used to determine the stretching stiffness of typical folded structures. Denoting $b(0)$ and $b(F)$ the folded structure lengths in the absence or presence of force, and assuming Hookean stretching elasticity, we have:

\begin{equation}\label{Eq_bF}
b(F)=b(0)+{b(0) \over \gamma}F,
\end{equation}
where $\gamma$ is the stretching stiffness and $b(0) \over \gamma$ describes the stretching deformability of the folded structure. Therefore, for a folded structure, $\gamma$ could be calculated from the linear dependence of $b(F)$ on F.

\begin{figure}[htbp]
\centering
\includegraphics[width=0.4\textwidth]{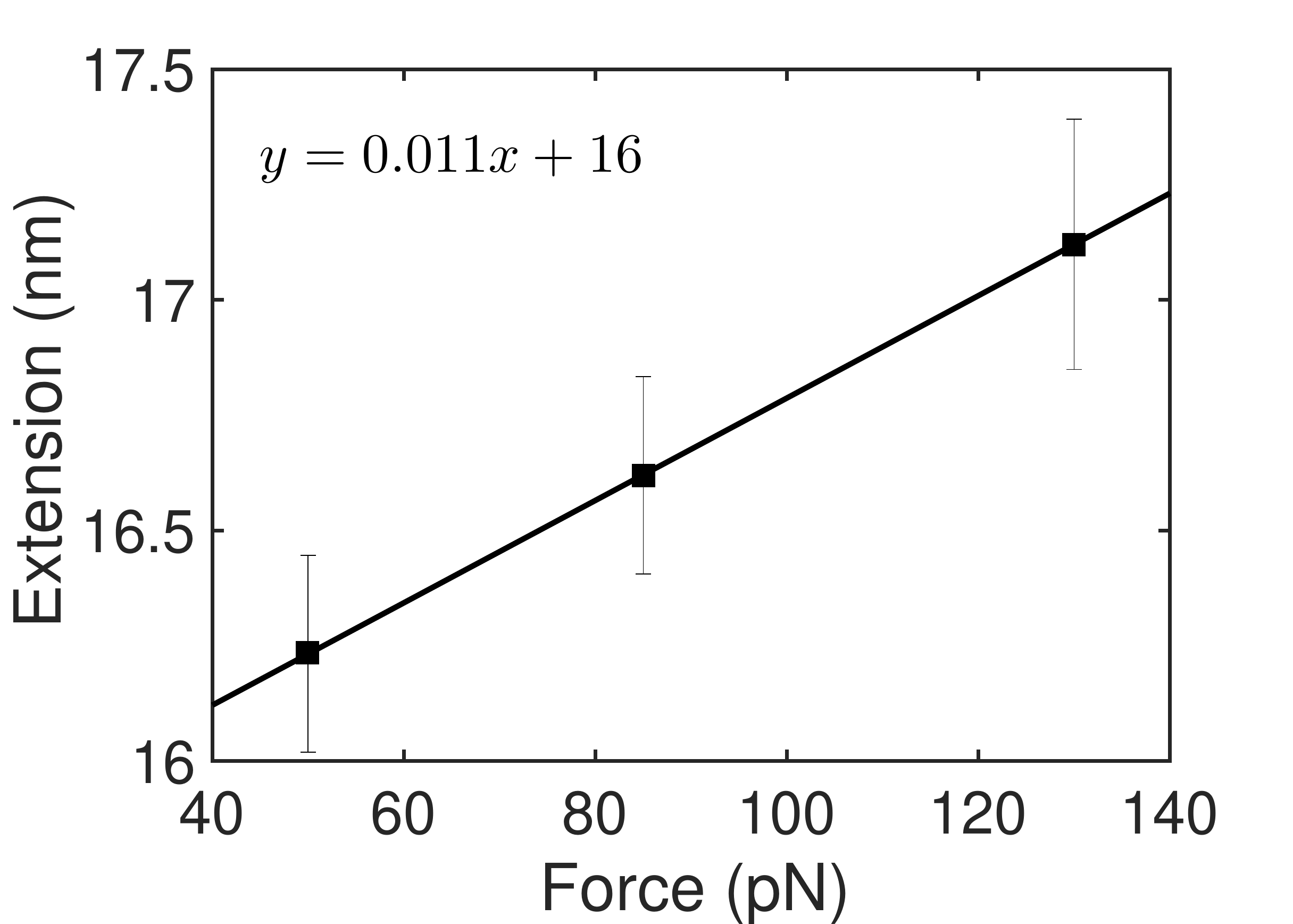}
\caption{{\bf Elasticity of dsDNA molecule.} Extension of 50 bp dsDNA was measured at constant forces (squares in the figure show the average value of extension and vertical error bars indicate the standard deviation). The simulations run for 50 ns before which no structural transition occurs, so the extensions were a pure elastic response. The fitting parameter of the slope $s$ is determined to be (with 95\% confidence bounds): $s=0.011~(0.011, 0.011)$ nm/pN. Based on the well-known B-form DNA contour length $\sim 0.34$ nm per basepair, the value of $b$ is determined as $b \sim 0.34 \times 50=17$ nm. As a result, the elasticity of dsDNA molecule is estimated to be $\gamma \sim 1500$ pN.}
\label{fig:dna_gamma}
\end{figure}

\begin{figure}[htbp]
\centering
\includegraphics[width=0.4\textwidth]{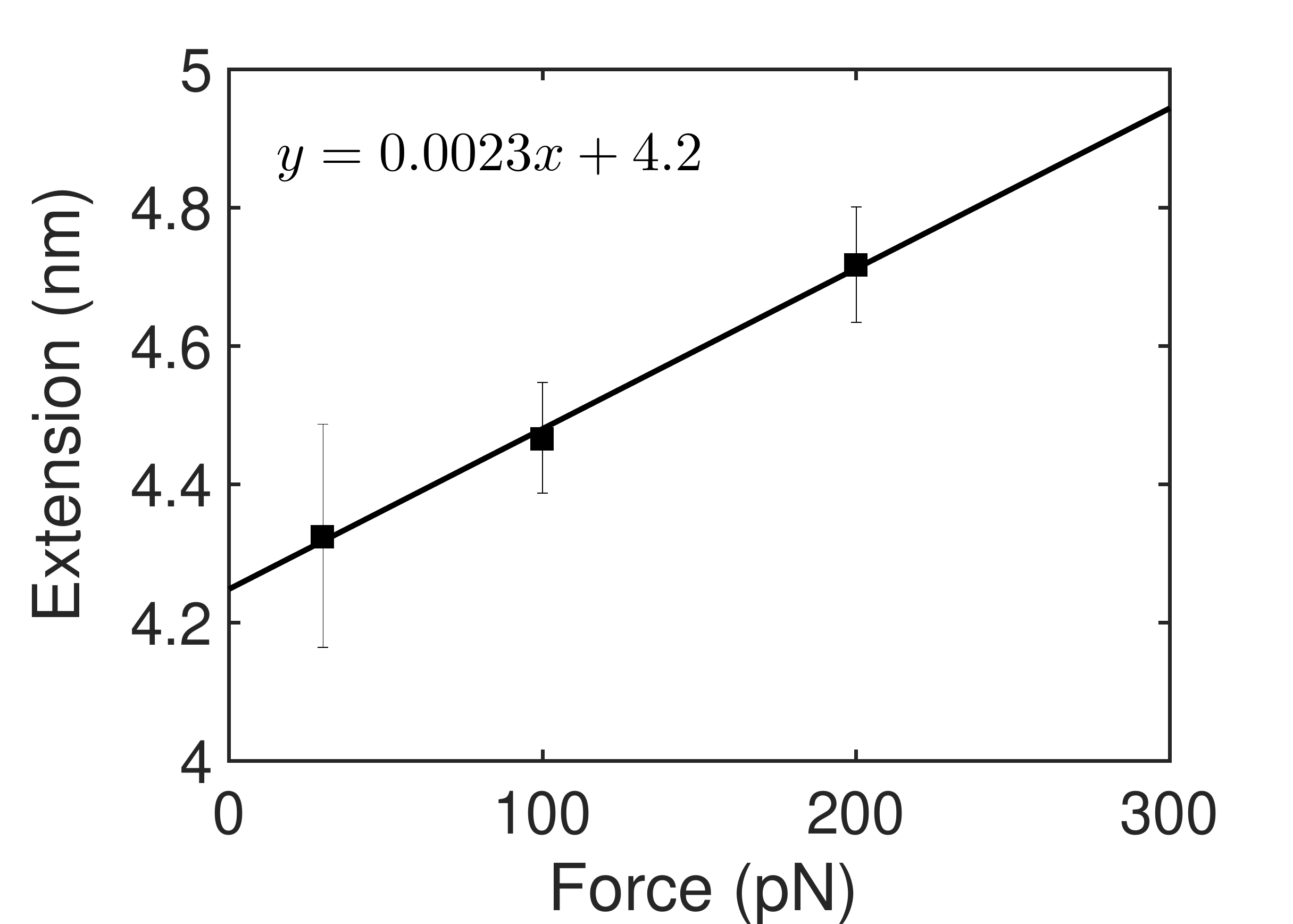}
\caption{{\bf Elasticity of titin I27 domain in native state.} Extension of titin I27 domain in the native state was measured at constant forces (squares in the figure show the average value of extension and vertical error bars indicate the standard deviation). The simulations run for 50 ns before which no structural transition occurs, so the extensions were a pure elastic response. The fitting parameter of the slope $s$ is determined to be (with 95\% confidence bounds): $s=0.0023~(0.0006, 0.0040)$ nm/pN. According to the value of $b \sim 4.3$ nm, which is obtained based on the structure of I27 in native state, the elasticity of folded titin I27 domain is estimated to be $\gamma \sim 1900$ pN. }
\label{fig:I27_gamma}
\end{figure}

\begin{figure}[htbp]
\centering
\includegraphics[width=0.4\textwidth]{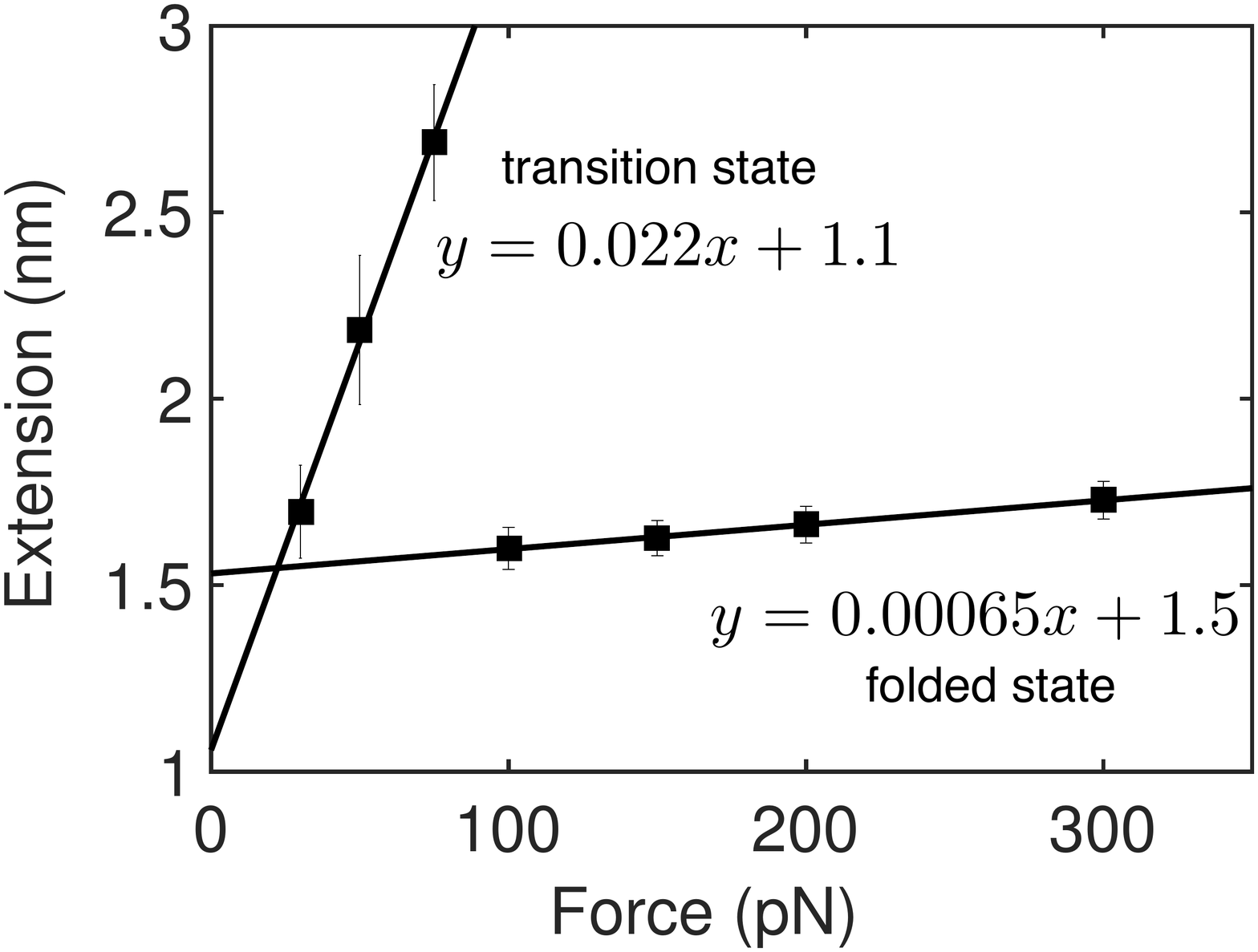}
\caption{{\bf Elasticity of src SH3 domain in native state and transition state.} Extension of src SH3 domain in the native and transition states were measured at constant forces (squares in the figure show the average value of extension and vertical error bars indicate the standard deviation). The simulations run for 50 ns before which no structural transition occurs, so the extensions were a pure elastic response. For the native state, the fitting parameter of the slope $s$ is determined to be (with 95\% confidence bounds): $s=0.00065~(0.00057, 0.00073)$ nm/pN. According to the value of $b \sim 1.90$ nm, which is obtained based on the structure of src SH3 in native state, the elasticity of  folded src SH3 is estimated to be $\gamma \sim 2900$ pN. Similarly,  the slope $s$ for transition state is determined to be $s=0.022~(0.007, 0.037)$ nm/pN. Since the folded core of the transition state maintains the overall structure as in the native state, the value of $b$ is also estimated to be $b \sim 1.90$ nm for the transition state, from which the elasticity of  the transition state of src SH3 is calculated to be $\gamma \sim 86$ pN.}
\label{fig:sh3_gamma}
\end{figure}

We calibrated this method for double-stranded DNA (dsDNA), whose stretching stiffness is in the range of $1000 - 3000$ pN, as measured from single-molecule stretching experiments \cite{bustamante2000single, Marko1995, Fu2011, Zhang2013}. The estimated $\gamma$ of dsDNA (150 mM NaCl) from our MD simulation is around $1500$ pN (Fig.~\ref{fig:dna_gamma}), which is consistent with experimentally measured values. Using this approach, we estimated $\gamma$ for the native state of titin I27 (Fig.~\ref{fig:I27_gamma}), as well as the native and transition states of src SH3 (Fig.~\ref{fig:sh3_gamma}).  

\begin{table}[htbp]
\centering
    \caption{$\gamma$ (pN) for different structures}
    \label{tab:gamma}
    \begin{tabular}{cccc}
    \hline
    dsDNA & I27 & SH3 native & SH3 transition \\
    \hline
    1500 & 1900 & 2900 & 86\\
    \hline
  \end{tabular}
 
\end{table}

\subsection*{S\rom{3}. The conformational free energy of a molecule stretched by an external constant force}
\begin{figure}[htbp]
\centering
\includegraphics[width=0.4\textwidth]{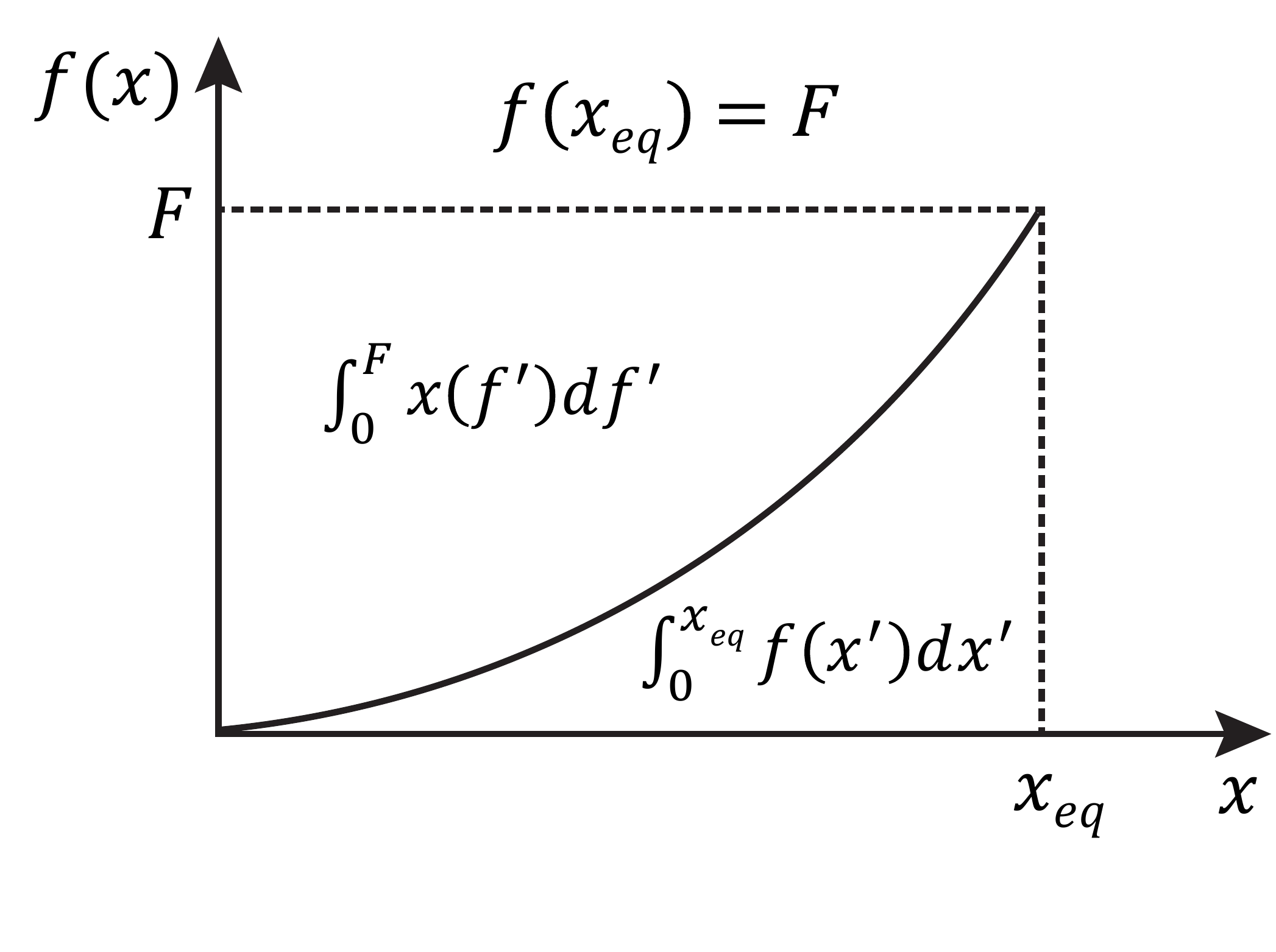}
\caption{{\bf The conformational free energy of a molecule under external force.} At equilibrium, the conformational free energy of a molecule under force $\Phi(x(F),F)=\int\limits_{0}^{x(F)} f(x') dx'-Fx(F)$ equals $\Phi(F)=\int\limits_{0}^F x(F') dF'$.}
\label{fig:Lgdr}
\end{figure}

In general, an external constant force $F$ applied to a molecule in a given state introduces a conformational free energy to the state by: 
\begin{equation}
\label{Eq_Cenergy1}
\Phi(x,F)=\int\limits_{0}^x f(x') dx'-Fx,
\end{equation}
where $x$ is the extension of the molecule in this state, and $F$ is the applied force. The external force contributes to a potential energy of $-Fx$. At equilibrium, $x$ is no longer independent from $F$, since it depends on $F$ through the force-extension curve $x(F)$. Therefore, this energy becomes dependent only on force: $\Phi(x(F),F)=\int\limits_{0}^{x(F)} f(x') dx'-Fx(F)$, which can be rewritten to a simpler form by Legendre transformation \cite{cocco2004overstretching, rouzina2001force}:
\begin{equation}
\label{Eq_Cenergy2}
\Phi(F)=\int\limits_{0}^F x(F') dF'.
\end{equation}
This can be easily seen by the relation (Fig:~\ref{fig:Lgdr}):
\begin{equation}
\label{Eq_Cenergy3}
\int\limits_{0}^{x_\text{eq}} f(x') dx'+\int\limits_{0}^F x(F') dF'=Fx_\text{eq}.
\end{equation}

\subsection*{S\rom{4}. Alternative derivation of Eq. 5}
Based on the force-dependent free energies of the molecule in both native and transition states that are shown in Eq. 3, $k(F)$ is determined by applying the Arrhenius law $k(F)  =k_\text{0} e^{-\beta \Delta \Phi^*(F)}$:
\begin{equation}
\label{Eq_k}
k(F)  =k_\text{0} e^{-\beta \left( \Phi_{b^*, \gamma^*}(F)-\Phi_{b^\text{0}, \gamma^\text{0}}(F)  +  \Phi_{{L^*}}(F)\right)}.
\end{equation}
In the main text, the large force expression Eq. 5 can be derived based on direct asymptotic expansion from Eq. 4. Here we provide an alternative derivation based on large-force expansion of force-extension curves of folded structure and flexible polymer. At large forces ($F \gg k_\text{B}T/b$ and $F \gg k_\text{B}T/A$), the force-extension curves of the extensible folded structure and the flexible polymer have very simple asymptotic expressions: 
\begin{equation}
\label{Eq_S_xF}
\left\{
\begin{array}{l}
x_{b,\gamma}(F)\approx b (1-{k_BT\over Fb})(1+{F\over \gamma}),\\
x_L(F)\approx L (1-\sqrt{k_\text{B}T \over 4A F}).\
\end{array}\right.
\end{equation}

These expressions are derived based on large force expansion ($F \gg k_\text{B}T/b$ and $F \gg k_\text{B}T/A$). The typical sizes of protein domain and the folded core in the transition state are in the order of a few nm; therefore, $k_\text{B}T/b$ are close to 1 pN. If in the transition state a protein peptide or a ssDNA/ssRNA polymer is produced, due to their very small bending persistence length of $A \sim 1$ nm, $k_\text{B}T/A \sim 5$ pN becomes the predominating factor that imposes a restriction to the lower boundary of force range. 

\begin{figure}[htbp]
\centering
\includegraphics[width=0.4\textwidth]{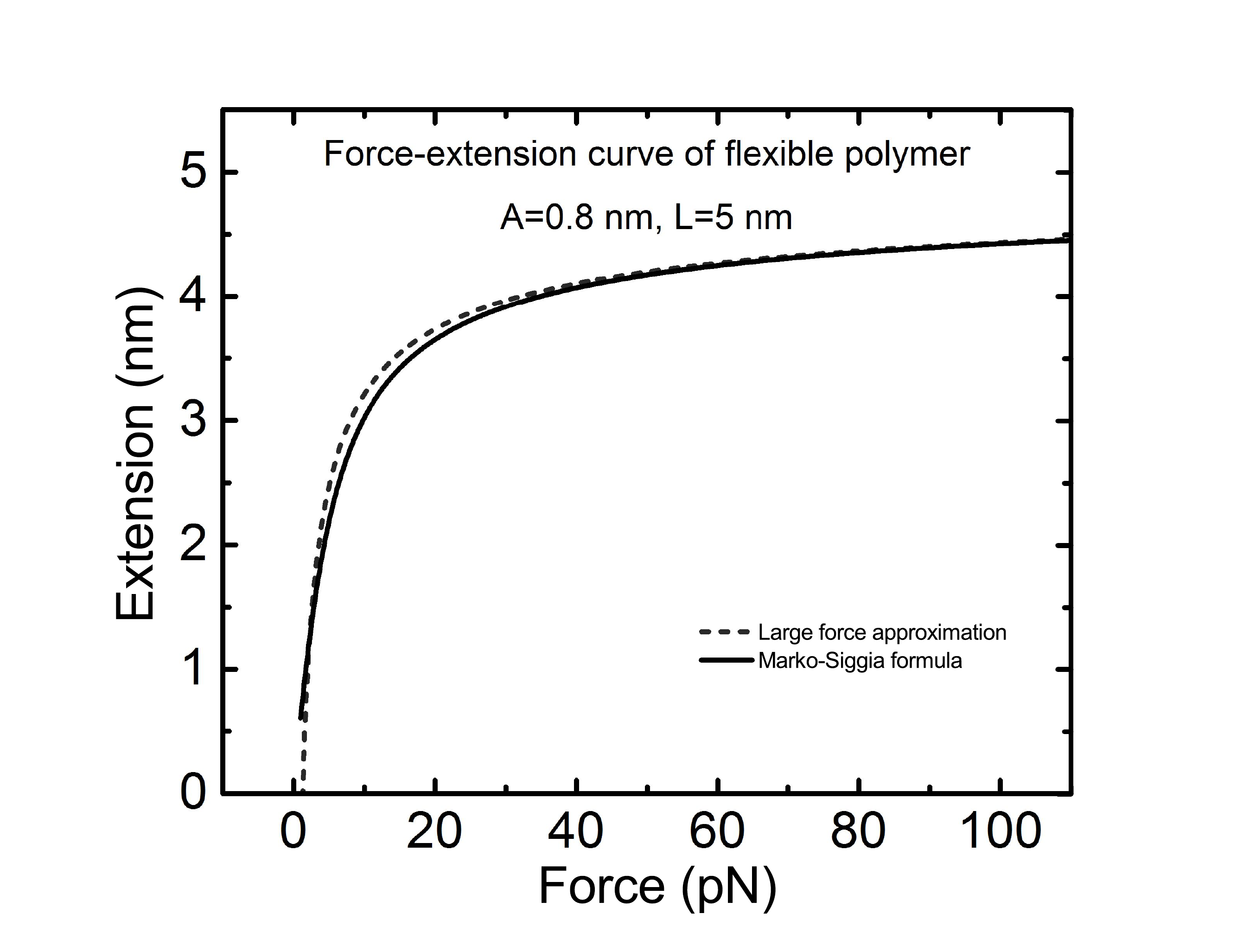}
\caption{{\bf Force-extension curves of flexible polymer.} The force-extension curve of a flexible polymer with $A=0.8$ nm and $L=5$ nm calculated based on the asymptotic large force expansion (dash line) differs from the one from the full Marko-Siggia formula (solid line) by less than 10\%.}
\label{fig:wlc}
\end{figure}

In actual applications, however, the applicable forces do not have to be much greater than 5 pN, since the force-extension curve of a flexible polymer with $A =0.8$ nm and $L=5$ nm calculated based on the asymptotic large force expansion differs from the one according to the full Marko-Siggia formula by less than 10\% (Fig.~\ref{fig:wlc}) at forces above 3 pN.

Based on these large-force asymptotic expressions of the force-extension curves, it is straightforward to show that the force-dependent change in the free energy barrier is:
\begin{equation}
\label{Eq_S_Phi_1}
\Delta \Phi^*(F) \approx -\left(\sigma F+ \alpha F^2/2 - \eta F^{1/2}\right).
\end{equation}

Here $\sigma = L^*+(b^*-b^\text{0})-({k_\text{B}T\over \gamma^*} - {k_\text{B}T \over \gamma^\text{0}})$, $\alpha={b^*\over \gamma^*} - {b^\text{0}\over \gamma^\text{0}}$, and $\eta=L^* \sqrt{k_\text{B}T\over A}$. Typical values of ${k_\text{B}T\over \gamma^\text{0}}$ and ${k_\text{B}T\over \gamma^*}$ are in the range of $10^{-3}$ nm  - $10^{-2}$ nm (SI: Sec \rom{2}); therefore, $\sigma \sim L^*+(b^*-b^\text{0})$. Eq. 5 is obtained by applying the Arrhenius law:
\begin{equation}
\label{Eq_S_k1}
k(F)  = \tilde{k}_\text{0} e^{\beta \left( \sigma F+ \alpha F^2/2 - \eta F^{1/2}\right) }.
\end{equation}

\subsection*{S\rom{5}. Bootstrap analysis to determine fitting errors} 
In order to test the robustness of fitting of Eq. 5 to experimental data, for each molecule studied in our work, we performed 1000 times of fitting with 80\% data points that are randomly chosen from the experimentally measured $k(F)$ data for every fitting. We found that all the 1000 sets of the best-fitting parameters are in the reasonable range around the best-fitting parameters that are determined using the whole experimental data. Table~\ref{tab:I27d}, Table~\ref{tab:PSGL1d} and Table~\ref{tab:SH3d} have shown the averages and the standard deviations of the best-fitting parameters ($\tilde{k}_\text{0}$, $\sigma$, $\alpha$, and $\eta$) for 1000 times of fitting with the randomly chosen data points, which occupy 80\% of the whole experimental data in each fitting. The structural-elastic parameters in the transition state determined based on the native state structure, steered MD simulation, or solved from the best-fitting parameters are also indicated in the tables.

\begin{table}[!htbp]
    \caption{Parameters for I27 by fitting Eq. 5 to 1000 sets of 80\% data points}
    \label{tab:I27d}
    \begin{tabular}{c|cccc|ccc}
    \hline
    &\multicolumn{4}{c|}{Best-fitting parameters} & \multicolumn{3}{c}{Structural-elastic parameters}\\
    & $\tilde{k}_\text{0}$ (s$^{-1}$) & $\sigma$ (nm) & $\alpha$ (nm/pN) & $\eta$ (nm$\cdot$pN$^{1/2}$) & $L^*$ (nm) & $b^*$ (nm) & $\gamma^*$ (pN) \\
    \hline
     Average & 0.026 & 1.014 & 0.003 & 10.023 & 4.4 & 0.9 & 179\\
     Standard deviation & 0.014 & 0.243 & 0.003 & 1.542 & 0.7 & 0.4 & 41\\ 
    \hline
  \end{tabular}
\end{table}

\begin{table}[!htbp]
    \caption{Parameters for sPSGL-1 / sP-selectin by fitting Eq. 5 to 1000 sets of 80\% data points}
    \label{tab:PSGL1d}
    \begin{tabular}{c|cccc|ccc}
    \hline
    &\multicolumn{4}{c|}{Best-fitting parameters} & \multicolumn{3}{c}{Structural-elastic parameters}\\
    & $\tilde{k}_\text{0}$ (s$^{-1}$) & $\sigma$ (nm) & $\alpha$ (nm/pN) & $\eta$ (nm$\cdot$pN$^{1/2}$) & $L^*$ (nm) & $b^*$ (nm) & $\gamma^\text{0}$=$\gamma^*$ (pN)\\
    \hline
     Average & 58.498 & 0.727 & -0.005 & 5.752 & 2.5 & 5.5 & 379\\ 
     Standard deviation & 27.083 & 0.162 & 0.001 & 1.275 & 0.6 & 0.4 & 48\\
    \hline
  \end{tabular}
\end{table}

\begin{table}[!htbp]
    \caption{Parameters for src SH3 by fitting Eq. 5 to 1000 sets of 80\% data points}
    \label{tab:SH3d}
    \begin{tabular}{c|ccc|cc}
    \hline
    &\multicolumn{3}{c|}{Best-fitting parameters} & \multicolumn{2}{c}{Structural-elastic parameters}\\
    & $\tilde{k}_\text{0}$ (s$^{-1}$) & $\sigma$ (nm) & $\alpha$ (nm/pN) & $b^*$ (nm) & $\gamma^*$ (pN)\\
    \hline
     Average & 0.045 & -0.475 & 0.050 & 1.4 & 29\\ 
     Standard deviation & 0.043 & 0.249 & 0.009 & 0.2 & 9\\  
    \hline
  \end{tabular}
\end{table}

\subsection*{S\rom{6}. Robustness of convergence of the best-fitting parameters} 
We have tested whether the best-fitting values of these parameters in Eq. 5 ($\sigma$, $\alpha$ and $\eta$) are uniquely determined for a given shape of $k(F)$, by starting from many well-separated different initial values for the fitting of $k(F)$.  We used lsqcurvefit function in Matlab to solve the nonlinear curve-fitting (data-fitting) problems in least-squares sense. By starting with different initial points for the fitting, lsqcurvefit may find a local solution that is not particularly close to the global best-fitting parameter values. So if another set of solutions exists that can fit equally well the data, one of the well-separated initial values may lead to a new set of solutions due to the existence of possible local minimums. However, for each of the three molecules tested in the study, we have found that the parameters always converge to the same set regardless of the initial values (Table \ref{tab:I27} for I27, Table \ref{tab:PSGL1} for sPSGL-1/ sP-selectin, Table \ref{tab:SH3} for src SH3), which means the best-fitting parameters can be uniquely determined when applying Eq. 5 to fit experimental data of $k(F)$, at least for all the three cases studied in this work. 

\begin{table}[!htbp]
    \caption{Best-fitting parameters for I27 with different initial values}
    \label{tab:I27}
    \begin{tabular}{|c|cccc|cccc|}
    \hline
    &\multicolumn{4}{c|}{Initial values} & \multicolumn{4}{c|}{Best-fitting values}\\
    Case & $\tilde{k}_\text{0}$ (s$^{-1}$) & $\sigma$ (nm) & $\alpha$ (nm/pN) & $\eta$ (nm$\cdot$pN$^{1/2}$) & $\tilde{k}_\text{0}$ (s$^{-1}$) & $\sigma$ (nm) & $\alpha$ (nm/pN) & $\eta$ (nm$\cdot$pN$^{1/2}$)\\
    \hline
     1 & 0.01 & -10 & -10 & 0 & 0.026 & 1.099 & 0.002 & 10.519\\
     2 & 0.01 & -10 & -10 & 5 & 0.026 & 1.099 & 0.002 & 10.519\\
     3 & 0.01 & -10 & -10 & 10 & 0.026 & 1.099 & 0.002 & 10.519\\
     4 & 0.01 & 0 & 0 & 5 & 0.026 & 1.099 & 0.002 & 10.519\\
     5 & 0.01 & 0 & 0 & 10 & 0.026 & 1.099 & 0.002 & 10.519\\
     6 & 0.01 & 10 & 0 & 5 & 0.026 & 1.099 & 0.002 & 10.519\\
     7 & 0.01 & 0 & 10 & 5 & 0.026 & 1.099 & 0.002 & 10.519\\
     8 & 1 & 0 & 0 & 10 & 0.026 & 1.099 & 0.002 & 10.519\\
     9 & 0.00001 & 0 & 0 & 10 & 0.026 & 1.099 & 0.002 & 10.519\\
    10 & 1 & 10 & 10 & 10 & 0.026 & 1.099 & 0.002 & 10.519\\
    \hline
  \end{tabular}
\end{table}
\begin{table}[!htbp]
    \caption{Best-fitting parameters for sPSGL-1/ sP-selectin with different initial values}
    \label{tab:PSGL1}
     \begin{tabular}{|c|cccc|cccc|}
    \hline
    &\multicolumn{4}{c|}{Initial values} & \multicolumn{4}{c|}{Best-fitting values}\\
    Case & $\tilde{k}_\text{0}$ (s$^{-1}$) & $\sigma$ (nm) & $\alpha$ (nm/pN) & $\eta$ (nm$\cdot$pN$^{1/2}$) & $\tilde{k}_\text{0}$ (s$^{-1}$) & $\sigma$ (nm) & $\alpha$ (nm/pN) & $\eta$ (nm$\cdot$pN$^{1/2}$)\\
    \hline
     1 & 1 & 0 & 0 & 0 & 51.786 & 0.723 & -0.005 & 5.760\\
     2 & 1 & 0 & 0 & 5 & 51.786 & 0.723 & -0.005 & 5.760\\
     3 & 1 & 0 & 0 & 10 & 51.786 & 0.723 & -0.005 & 5.760\\
     4 & 1 & 5 & 0 & 0 & 51.786 & 0.723 & -0.005 & 5.760\\
     5 & 1 & 10 & 0 & 0 & 51.786 & 0.723 & -0.005 & 5.760\\ 
     6 & 1 & 10 & 10 & 10 & 51.786 & 0.723 & -0.005 & 5.760\\ 
     7 & 1 & 0 & -5 & 5 & 51.786 & 0.723 & -0.005 & 5.760\\
     8 & 1 & 0 & -10 & 5 & 51.786 & 0.723 & -0.005 & 5.760\\
     9 & 10 & 0 & 0 & 5 & 51.786 & 0.723 & -0.005 & 5.760\\
    10 & 100 & 0 & 0 & 5 & 51.786 & 0.723 & -0.005 & 5.760\\
    \hline
  \end{tabular}
\end{table}
\begin{table}[!htbp]
    \caption{Best-fitting parameters for src SH3 with different initial values}
    \label{tab:SH3}
    \begin{tabular}{|c|ccc|ccc|}
    \hline
    &\multicolumn{3}{c|}{Initial values} & \multicolumn{3}{c|}{Best-fitting values}\\
    Case & $\tilde{k}_\text{0}$ (s$^{-1}$) & $\sigma$ (nm) & $\alpha$ (nm/pN) & $\tilde{k}_\text{0}$ (s$^{-1}$) & $\sigma$ (nm) & $\alpha$ (nm/pN)\\
    \hline
     1 & 0.1 & 0 & 0 & 0.030 & -0.441 & 0.049\\
     2 & 0.1 & 0 & 10 & 0.030 & -0.441 & 0.049\\
     3 & 0.1 & 0 & 100 & 0.030 & -0.441 & 0.049\\
     4 & 0.1 & 10 & -10 & 0.030 & -0.441 & 0.049\\
     5 & 0.1 & 10 & -100 & 0.030 & -0.441 & 0.049\\
     6 & 0.1 & -10 & 10 & 0.030 & -0.441 & 0.049\\
     7 & 0.1 & -100 & 10 & 0.030 & -0.441 & 0.049\\
     8 & 10 & 10 & 10 & 0.030 & -0.441 & 0.049\\
     9 & 100 & 10 & 10 & 0.030 & -0.441 & 0.049\\
    10 & 100 & 100 & 100 & 0.030 & -0.441 & 0.049\\
    \hline
  \end{tabular}
\end{table}

\subsection*{S\rom{7}. Fitting of Eq. 5 to experimental data of src SH3} 
\begin{figure}[htbp]
\centering
\includegraphics[width=0.4\textwidth]{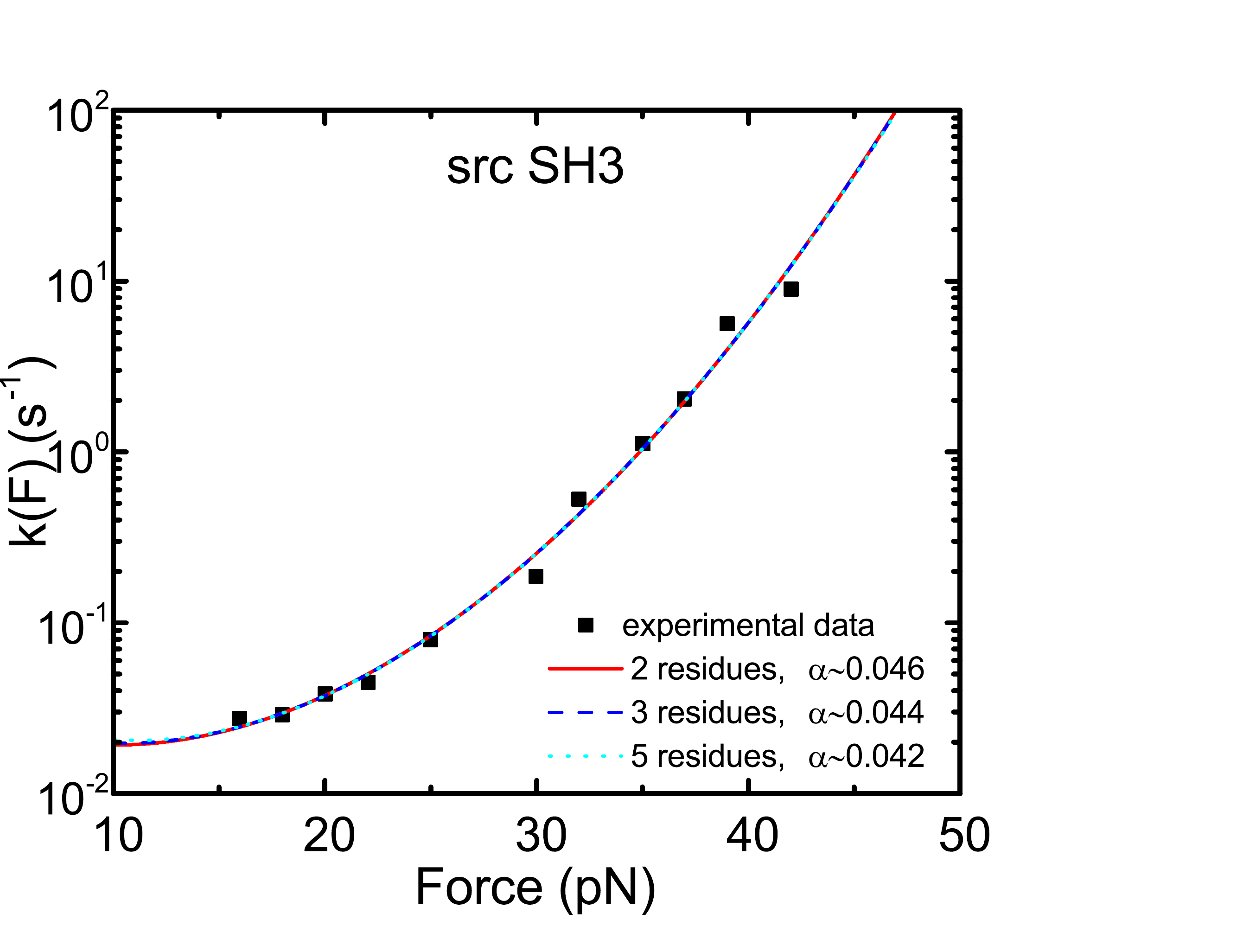}
\caption{{\bf Fitting of Eq. 5 to experimental data of src SH3.} By restricting the number of residues of the flexible peptide in the transition state of src SH3, good quality of fitting can be obtained. The figure shows the fitting curves of experimental data for src SH3 domain when the number of peptide residue is presupposed to be 2, 3 and 5. The goodness-of-fit is evaluated by R-square $\sim 0.992$ for all the fittings.}
\label{fig:sh3_diff_rsd}
\end{figure}

In the fitting of Eq.5 to the $k(F)$ data of src SH3, $\eta < 4.3 $ nm$\cdot$pN$^{1/2}$ is needed to ensure a positive $b^*$. By restricting the number of residues of the flexible peptide in the transition state of src SH3, good quality of fitting can be obtained (Fig.~\ref{fig:sh3_diff_rsd}), which suggests that the peptide length is not a key factor for the $k(F)$ profile. At each peptide length, the best-fitting parameters predict $\alpha > 0$ and $\gamma^* \ll \gamma^\text{0}$, indicating that a much softer folded core in the transition state than that of the native state is the predominant factor of $k(F)$ (Table \ref{tab:sh3 fitting}). Previous study estimated a small transition distance $\sim 0.45$ nm in the force range of $15-25$ pN \cite{Jagannathan2012}, suggesting insignificant fraction of peptide in the transition state. Otherwise, considering $0.22 - 0.28$ nm per residue of typical peptide in $15-25$ pN force range \cite{Winardhi2016}, one would expect a significantly larger transition distance if a long peptide ($> 3$ a.a) is produced in the transition state. 

\begin{table}[htbp]
    \caption{Fitting parameters for src SH3}
    \label{tab:sh3 fitting}
    \begin{tabular}{cccccc}
    \hline
    $n^*$ & $L^*$ (nm) & $\eta$ (nm$\cdot$pN$^{1/2}$) & $\sigma$ (nm) & $\alpha$ (nm/pN) & $\gamma^*$ (pN) \\
    \hline
     1 & 0.38 & 0.86 & -0.317 & 0.048 & 25\\
     2 & 0.76 & 1.7 & -0.196 & 0.046 & 20\\
    3 & 1.14 & 2.6 & -0.066 & 0.044 & 16\\
    4 & 1.52 & 3.4 & 0.049 & 0.043 & 10\\
    5 & 1.90 & 4.3 & 0.179 & 0.042 & 4\\
    \hline
  \end{tabular}
  
{$n^*$ is the number of residues assumed for the peptide length in the transition state of src SH3. $L^*$ is the contour length of the flexible polymer, which is determined based on $L^*=n^* l_\text{r}$ and $l_\text{r}\sim 0.38$ nm for peptide chain. The value of $\eta$ is restricted by $\eta=L^* \sqrt{k_\text{B}T\over A}$ in the fitting of Eq. 2 to the experimental data of src SH3 for each peptide length. $\sigma$ and $\alpha$ are the besting fitting values. Based on the structure of the native state and using steered MD simulation, the structural-elastic parameters of the native state are determined to be $b^\text{0} \sim 1.90$ nm and $\gamma^\text{0} \sim 2900$ pN (S\rom{2}). From these parameters, the value of $\gamma^*$ was solved for each presupposed peptide length. The goodness-of-fit is evaluated by R-square $\sim 0.992$ for all the fittings.}
\end{table}

\clearpage

\subsection*{Supplementary figures}

\end{document}